\documentclass[prd,11pt]{revtex4-1}
\usepackage{graphicx}
\usepackage{amsmath}
\usepackage{amssymb}
\usepackage{bm}
\usepackage{array}
\usepackage{booktabs}
\usepackage{siunitx}
\usepackage{multirow}
\usepackage{subcaption}
\usepackage{float}
\usepackage{hyperref}
\usepackage{slashed}
\usepackage{amsfonts}
\usepackage{xcolor}
\usepackage{mathrsfs}
\usepackage{CJK}

\hypersetup{
    colorlinks=true,
    linkcolor=blue,
    filecolor=magenta,
    urlcolor=cyan,
    citecolor=green!40!black,  
}
\usepackage[utf8]{inputenc}
\usepackage[T1]{fontenc}
\usepackage{lipsum} 

\begin{document}

\title{Distribution Amplitudes of Pseudoscalar Mesons within the Light-Front Quark Model}

\author{Xiao-Nan Li$^{a}$}
\author{Shuai Xu$^{b}$}
\email{sxu_iopp@foxmail.com}
\author{Qin Chang$^{c}$}

\affiliation{
$^a$School of Electrical Engineering, Tongling University, Tongling 244061, P.R. China\\
$^b$School of Physics and Telecommunications Engineering, Zhoukou Normal University, Zhoukou 466001, P.R. China\\
$^c$Institute of Particle and Nuclear Physics, Henan Normal University, Xinxiang 453007, China}

\date{\today}

\begin{abstract}	
In this paper, we investigate the distribution amplitudes (DAs) of pseudoscalar mesons within the light-front quark model (LFQM), and the $n$-th Gegenbauer moment $a_n(\mu)$, $\xi$-moment $\langle \xi^n\rangle$ and transverse moment $\langle \mathbf{k}^n_\perp\rangle$ are also studied. Two parameter sets are adopted by fitting to mesonic decay constants or mass spectra, leading to distinct schemes for numerical analysis. Our results reveal several key insights: (1) Under scheme-I, the second Gegenbauer moment for the pion, $a_{2,\pi} = 0.054$, is in excellent agreement with predictions from the platykurtic model and the nonlocal chiral quark model (NL$\chi$QM). Meanwhile, scheme-II yields $a_{2,\pi} = 0.126$, which is consistent with lattice QCD (LQCD), Dyson-Schwinger equation (DSE) and data-driven light-cone sum rules (LCSR) analyses. Furthermore, we observe that $\phi_{2,\pi} = \phi^P_{3,\pi}$ with the replacement $M \to M_0$. (2) Flavor symmetry breaking effects are found to be more significant in higher-twist DAs, with a notable relation $a_1^P\approx 2 a_1$ emerging in heavy meson systems. (3) Interestingly, numerical analysis indicates that the transverse moment $\langle \mathbf{k}^n_\perp\rangle$ is twist-independent with the replacement $M\to M_0$. Additionally, we propose an empirical scaling relation $\frac{\Delta m^{0.61}}{\beta^{1.10}}x_p=0.67$ to effectively describe the shape of DAs for heavy mesons. Collectively, these results demonstrate that mass asymmetry remarkably influences DAs and kinds of moments, these theoretical predictions will be tested in future experiments.
\end{abstract}

\maketitle

\section{Introduction}

A fundamental challenge in hadron physics is elucidating the internal structure of hadrons in terms of their constituent quarks and gluons. The distribution amplitudes (DAs) of mesons serve as essential quantities in this endeavor, describing the spatial distribution of partons in terms of their longitudinal momentum fractions $x=p^+/P^+$ inside the bound state~\cite{Lepage:1980fj,Brodsky:1997de,Efremov:1979qk,Collins:1981uw}. These nonperturbative quantities provide crucial information about mesonic structure and the dynamics governing constituent interactions~\cite{Braun:1989iv}. It follows naturally that many nontrivial mesonic properties, such as decay constants and various form factors, are mutually influenced by and intertwined with the DAs~\cite{Jaus:1989au,Jaus:1989av,Cheng:1996if,ODonnell:1996sya,Cheung:1996qt,Choi:1996mq,Barik:1997qq,Hwang:2000ez,Hwang:2010hw,Geng:2001de,Chang:2018aut}. Moreover, according to the factorization theorem and the constraints from chiral symmetry, the DAs (at least at the leading-twist level) exhibit universality in various decay modes, which has been confirmed by both experiments and LQCD simulations. Given these, DAs play a critical role and serve as indispensable inputs in phenomenological researches. For instance, the refined calculations for $B$-physics under the QCD factorization framework are highly relevant to the details of the DAs and their derivative physical quantities~\cite{Chang:2016ouf,Chang:2018mva,Chang:2019obq,Chang:2019mmh,Choi:2017uos}. For hard processes with a large momentum transfer, perturbative QCD (pQCD) factorizes the physical quantity into a convolution of the hard scattering kernel and the DAs~\cite{Efremov:2009dx,Collins:1996fb}. Due to the power suppression of $\mathcal{O}(\Lambda_{QCD}/Q)$, the higher twist structures provide small contributions (less than $5\%$)~\cite{Dong:2008zza} at high energy scales $\mu\gg \Lambda_{QCD}$ and then the leading-twist-2 and next-leading-twist-3 DAs offer a straightforward and powerful probe for achieving this goal.

Theoretically, numerous approaches have been applied to compute these quantities, including LQCD~\cite{LatticeParton:2022zqc,Chen:2023byr,RQCD:2019osh}, QCD Sum Rules (QCDSR)~\cite{Braun:1989iv,Chernyak:1983ej,Huang:2013yya,Khodjamirian:2004ga,Ball:2006wn,Guo:1991eb}, pQCD~\cite{Chai:2025xuz,Cheng:2019ruz}, DSE approach~\cite{Chang:2013pq,Raya:2015gva,Chang:2013epa,Shi:2015esa,Roberts:2021nhw}, the NL$\chi$QM~\cite{Nam:2006sx,Nam:2006au}, relativistic LFQM~\cite{Hwang:2010hw,Choi:2017uos}, LCSR~\cite{Agaev:2010aq,Agaev:2012tm,Cheng:2020vwr,Chai:2022srx} and others~\cite{Petrov:1998kg,Nam:2006mb,RuizArriola:2002bp,Praszalowicz:2001wy}. Leveraging the advantages of manifest Lorentz invariance and conceptual simplicity, the LFQM has been quite successfully used in analyzing various hadronic form factors, decay constants, distribution amplitudes, etc~\cite{Choi:2009ai,Zhao:2018zcb,Wuenqi:2025tup,Xu:2025aow,Ke:2010vn,Xu:2025ntz}. It is found that the LFQM permits a reliable prediction of data on the electroweak transitions of pseudoscalar and vector mesons. In this work, we utilize the LFQM framework to study the DAs of pseudoscalar mesons.

In particular, the light-front dynamics (LFD) carries the maximum number (seven) of the kinematic (or interaction-independent) generators and thus less effort in dynamics is necessary in order to get the QCD solutions that reflect the full Poincaré symmetries~\cite{Dirac:1949cp,Brodsky:1998hn,Keister:1991sb,Coester:1992cg,Szczepaniak:1995vn,Osborn:1972dy,Susskind:1967rg,Bardakci:1968zqb,Chang:1968bh,Brodsky:1973kb,Bjorken:1970ah,Neville:1971uc}. Moreover, the rational energy momentum dispersion relation of LFD, namely $p^{-}=(\mathbf{p}_{\perp}^{2}+m^{2})/p^{+}$, yields the sign correlation between the light-front (LF) energy and the LF longitudinal momentum and leads to the suppression of quantum fluctuations of the vacuum, sweeping the complicated vacuum fluctuations into the zero modes in the limit of $p^+\rightarrow0$~\cite{Choi:2017uos}. This simplification is a remarkable advantage in LFD and facilitates the partonic interpretation of the amplitudes. The basic ingredient in LF QCD is the relativistic hadron wave function. It contains principally all information of a bound state in principle and generalizes the DAs by integrating the transverse momentum. The hadronic quantities are represented by the overlap of wave functions and can be derived naturally. In the LFQM, hadrons are composed of valence quarks and the equation of motion of the bound $q\bar{q}$ meson in the LF formalism is a relativistic Schr\"{o}dinger equation with an effective confining potential, e.g., $\mathcal{H}_{LF}|\Psi\rangle=M|\Psi\rangle$, where the LF Hamiltonian of the bound state $\mathcal{H}_{LF}=H_0+V_{q\bar{q}}$ ~\cite{Brodsky:2014yha} and $|\Psi\rangle$ contains radial and spin components. Generally, the former wave function $\psi(x,\mathbf{k}_\perp)$ is obtained by solving the LF equation, while a fully relativistic treatment of the quark spins and the center-of-mass motion is implemented via the Melosh rotation~\cite{Melosh:1974cu,Cheng:1996if,Hwang:2010hw}. However, except in some simple cases, the full solution has remained a challenge. Rather than calculating these wave functions from a phenomenological potential, one often starts with an empirical trial wave function. There are several popular phenomenological wave functions that have been suggested in the literature.
In this work, a widely used Gaussian-type wave function is adopted to describe the bound state. The specific form of this wave function is determined by the masses of the constituent quarks and the parameter $\beta_{q\bar{q}}$. Among these, $\beta_{q\bar{q}}$ is a crucial parameter that essentially determines the confinement scale of the strong interaction, thereby directly relating to the spatial size of the bound state.
 The two parameters of the model can be fixed by a fit to the data of mesonic decay constants or masses. In our previous works, we have explored the quality and power of the LFQM in the studies of mesonic decay constants and kinds of form factors.
The main purpose of this paper is to study the DAs of pseudoscalar states within the LFQM. At the moment, we present three key physical moments: the Gegenbauer moment $a_n(\mu)$, which describes the degree of deviation of the DAs from their asymptotic form; the $\xi$-moment $\langle \xi^n \rangle$, which characterizes the longitudinal momentum distribution of the quarks; and the transverse moment $\langle k_\perp^n \rangle$, which reflects the transverse scale of the bound state. These results provide important information for a deeper understanding of the internal structure of pseudoscalar mesons.

The paper is organized as follows: Section~\ref{sec:2} provides a brief review of the theoretical formalism for the LFQM and the DAs. Numerical results and discussion are presented in Section~\ref{sec:3}. Finally, conclusions are given in Section~\ref{sec:4}.

\section{Theoretical Formulation} \label{sec:2}
\subsection{Briefly review for light-front quark model}\label{sec:model describtion}
Based on the LFD, the core concept of LFQM is the Fock state expansion of the $q\bar{q}$ bound state with the total momentum $P=(P^+,P^-,\mathbf{P}_\perp)$, where the Fock state is tread as in a noninteraction $q\bar{q}$ representation and the interaction are encoded in the LF wave function $\Psi_{h\bar{h}}(x,\mathbf{k}_\perp)$. Specifically, the expansion for a meson is given by
\begin{eqnarray}
|M(P)\rangle = \int \{d^3p_q\}\{d^3p_{\bar{q}}\}2(2\pi)^3\delta^3(P-p_q-p_{\bar{q}}) \sum_{h,\bar{h}}\Psi_{h\bar{h}}(x,\mathbf{k}_\perp)|q(p_q,h)\bar{q}(p_{\bar{q}},\bar{h})\rangle,
\label{eq:1}
\end{eqnarray}
where $p_q (p_{\bar{q}})$ and $h (\bar{h})$ is momentum and helicity of the constituent quark (antiquark), respectively. The momentum assignments in terms of LF variable $(x,\mathbf{k}_\perp)$ for constituents are as follows
\begin{eqnarray}
p^+_q &=& x P^+,~~~~~~~ p^+_{\bar{q}}=\bar{x} P^+,\\
\mathbf{p}_{q \perp} &=& x\mathbf{P}_\perp+\mathbf{k}_\perp,~~ \mathbf{p}_{\bar{q} \perp} =\bar{x}\mathbf{P}_\perp-\mathbf{k}_\perp,
\label{eq:3}
\end{eqnarray}
where $p_q^+$ and $\mathbf{p}_{q \perp}$ are longitudinal and transverse momentum. Inwhich, $\bar{x}=1-x$ and $P=p_q+p_{\bar{q}}$ is inherently satisfied. It is reasonable to set the $\mathbf{P}_\perp=0$ in LF frame of reference and then the transverse momentum for $q(\bar{q})$ is simplified to $\mathbf{p}_{q \perp}=\mathbf{k}_\perp$ and $ \mathbf{p}_{\bar{q} \perp}=-\mathbf{k}_\perp$. The LF wave function $\Psi_{h\bar{h}}(x,\mathbf{k}_\perp)$ is generally defined as
\begin{eqnarray}
\Psi_{h\bar{h}}(x,\mathbf{k}_\perp) = \psi(x,\mathbf{k}_\perp)S_{h\bar{h}}(x,\mathbf{k}_\perp),
\label{eq:4}
\end{eqnarray}
where $\psi(x,\mathbf{k}_\perp)$ is the radial wave function and $S_{h\bar{h}}(x,\mathbf{k}_\perp)$ corresponds to the spin wave function.
For pseudoscalar mesons, a kinds of 1S state radial wave function $\psi(x,\mathbf{k}_\perp)$ have been suggested in previous works\cite{Chang:2019mmh,Choi:2017uos} and the well-proven Gaussian-type wave function is taken in this work
\begin{eqnarray}
\psi(x,{\bf{k}}_{\perp})= \frac{4\pi^{3/4}}{\beta^{3/2}}\sqrt{\frac{\partial k_z}{\partial x}} \mathrm{exp}(-\frac{{\bf{k}}_{\perp}^2+k_z^2}{2\beta^2}), \label{eq:5}
\end{eqnarray}
where the parameter $\beta$ is meson-dependent and related to the size of bound state. With the confinements from mesonic decay constants or mass spectrum, the values of $\beta$ could be well fixed and the details are presented in the next section. The Jacobian $\frac{\partial k_z}{\partial x}$ originates from variable transformation $\{x,{\bf{k}}_{\perp}\}\rightarrow \vec{k}\equiv\{{\bf{k}}_{\perp},k_z\}$ and the detail is
\begin{eqnarray}
\frac{\partial k_z}{\partial x}= \frac{M_0}{4x\bar{x}}\Large\{1-\big[\frac{(m_q-m_{\bar{q}})^2}{M_0^2}\big]^2\Large\}, \label{eq:6}
\end{eqnarray}
where the longitudinal momentum $k_z=(x-\frac{1}{2})M_0+\frac{m_q^2-m_{\bar{q}}^2}{2M_0}$. The invariant mass of bound state $M_0$ is defined as
\begin{eqnarray}
M_0^2= \frac{m_{q}^2+{\bf{k}}_{\perp}^2}{x}+\frac{m_{\bar{q}}^2+{\bf{k}}_{\perp}^2}{\bar{x}},
\label{eq:7}
\end{eqnarray}
which plays a significant role in analyses of covariance and self-consistency of the standard and covariant LFQM~\cite{Chang:2019mmh,Chang:2019obq,Choi:2017uos}. It is easy to check that $M_0^2=(p_q+p_{\bar{q}})^2$ and  $M_0=E_q+E_{\bar{q}}$ inwhich the kinetic energy of constituents is $E_{q({\bar{q}})}=\sqrt{m_{q({\bar{q}})}^2+\vec{k}^2}$.

For LF spin wave function, $S_{h\bar{h}}(x,\mathbf{k}_\perp)$ is obtained by the interaction independent Melosh transformation from the traditional spin-orbit wave function and the covariant form is given by
\begin{eqnarray}
S_{h\bar{h}}(x,\mathbf{k}_\perp)= \frac{\bar{u}(p_q,h)\gamma_5 \nu(p_{\bar{q}},h)}{\sqrt{2}\sqrt{M_0^2-(m_q-m_{\bar{q}})^2}},
\label{eq:8}
\end{eqnarray}
where $u$, $v$ are Dirac spinor and $\sum_{h,\bar{h}}S^\dagger S=1$.

The formalism for bound state expansion has now been essentially clarified. When computing matrix elements, detailed expressions of the quark fields are frequently required, which will be presented below. The one particle state is defined by
\begin{eqnarray}
|q(p_q,h)\rangle=\sqrt{2p_q^+}b^\dagger(p_q)|0\rangle,~~~|\bar{q}(p_{\bar{q}},\bar{h})\rangle=\sqrt{2p_{\bar{q}}^+}d^\dagger(p_{\bar{q}})|0\rangle
\label{eq:9}
\end{eqnarray}
where the $b^\dagger(d^\dagger)$ is the creation operator of the valence quark (antiquark) in the bound state. The anticommutation relations for equal LF time $\tau=t-z=0$ are as follows
\begin{eqnarray}
\{b^\dagger(p_1),b(p_2)\}_{\tau=0}=(2\pi)^3\delta(p_1^+-p_2^+)\delta^2(\mathbf{p}_{1\perp}-\mathbf{p}_{2\perp})\delta_{h\bar{h}},
\label{eq:10}
\end{eqnarray}
so do as the one $\{d^\dagger(p),d(p)\}$. The quark field could be expanded in terms of creation and annihilation operator as
\begin{eqnarray}
q(x)=\int\frac{dp_q^+}{\sqrt{2p_q^+}}\frac{d^2\mathbf{p}_{q\perp}}{(2\pi)^3}\sum_h[b_h(p_q^+,\mathbf{p}_{q\perp})u_h(p_q^+,\mathbf{p}_{q\perp})
e^{-ip_q\cdot x}+d_h^\dagger(p_q^+,\mathbf{p}_{q\perp})\nu_h(p_q^+,\mathbf{p}_{q\perp})e^{ip_q\cdot x}]
\label{eq:11}
\end{eqnarray}
and the expression for $\bar{q}(x)$ can be obtained by taking the conjugate of Eq.~\ref{eq:11}. With the expressions of meson and quark (antiquark) field in Eqs.~\ref{eq:1},\ref{eq:11}, the matrix elements for $\langle 0|\bar{q}(z)\Gamma q(-z)|M(P)\rangle$ are derived as
\begin{eqnarray}
\mathcal{M}_{\Gamma}=\sqrt{N_c}\sum_{h,\bar{h}}\int\frac{dp_q^+}{\sqrt{2p_q^+}}\frac{d^2\mathbf{p}_\perp}{(2\pi)^3}
\Psi_{h,\bar{h}}(x,\mathbf{k}_\perp)\bar{\nu}_{\bar{h}}\Gamma u_h e^{i(2x-1)P\cdot z}.
\label{eq:12}
\end{eqnarray}
With the above theoretical preparations, physical quantities can be obtained through the computation of matrix elements. It is ready to give the DAs and relative kinds of moment in LFQM.
\subsection{DAs and kinds of moments}\label{sec:DAs}
Specifically, the leading twist DAs describe the probability amplitudes to find the hadron in a Fock-state with the minimum number of constituents, and the higher-twist DAs may come from the contributions of the higher Fock-state. In the framework of LFQM, the DAs of mesons are defined in terms of the following matrix elements
\begin{eqnarray}
\langle 0|\bar{q}(z)\gamma^\mu\gamma^5 q(-z)|M(P)\rangle = if_M P^\mu\int_0^1 dx~e^{i\xi P\cdot z}\phi_{2}(x,\mu),
\label{eq:13}
\end{eqnarray}
for the leading twist-2 DA $\phi_2(x,\mu)$ and
\begin{eqnarray}
\langle 0|\bar{q}(z)i\gamma^5 q(-z)|M(P)\rangle &=& f_M\mu_M\int_0^1 dx~e^{i\xi P\cdot z}\phi_{3}^P(x,\mu),
\label{eq:14}
\end{eqnarray}
for the next leading twist-3 pseudoscalar DA $\phi_{3,M}^P(x,\mu)$, respectively. $P$ denotes the 4-momentum of the meson and the path-ordered gauge link for the gluon fields between the point $-z$ and $z$ is taken. The integration variable $x$ is the longitudinal momentum fraction of the quark and $\xi=x-\bar{x}=2x-1$ depicts longitudinal separation. The $f_M$ is the decay constant of the meson and the $\mu_M=M^2/(m_q+m_{\bar{q}})$ is normalization factor which will be discussed below with the replacement of $M\to M_0$.
A few mathematical techniques are necessary to extract the DAs from Eqs.~\ref{eq:13}-\ref{eq:14}. By taking the Fourier transform for Eq.~\ref{eq:13} with the redefined variable $z^{\mu}=\tau\eta^{\mu}$ where the lightlike vector $\eta=(1,0,0,-1)$, we can obtain
\begin{eqnarray}
\int_{-\infty}^\infty d\tau \langle 0|\bar{q}(\tau\eta)\gamma^\mu\gamma^5 q(-\tau\eta)|M(P)\rangle e^{-i\xi'\tau P\cdot \eta}=if_M P^\mu\int_{-\infty}^\infty d\tau \int_0^1 dx~\phi_{2}(x,\mu)e^{i(\xi-\xi')\tau P\cdot \eta},
\label{eq:15}
\end{eqnarray}
inwhich $\xi'=2x'-1$ with dummy variable $x'$ corresponds to the conjugate variables for $\tau$. Substituting Eq.~\ref{eq:12} into left hand side of Eq.~\ref{eq:15}, we can directly extract the DA $\phi_{2}(x,\mu)$ as
\begin{eqnarray}
\phi_{2}(x,\mu)=\frac{1}{if_M P^\mu}\sqrt{N_c}\sum_{h,\bar{h}}\int\frac{dp_q^+}{\sqrt{2p_q^+}}\frac{d^2\mathbf{p}_\perp}{(2\pi)^3}
\Psi_{h,\bar{h}}(x,\mathbf{k}_\perp)\bar{\nu}_{\bar{h}}\gamma^\mu\gamma^5 u_h.
\label{eq:16}
\end{eqnarray}
Applying the LF wave function and performing the spinor contraction with $\bar{u}u=\slashed{p}$, we simplify the right hand side of Eq.~\ref{eq:16}
\begin{eqnarray}
\phi_{2}(x,\mu)=\frac{\sqrt{N_c}}{if_M P^\mu}\int^{\mu^2}\frac{d^2\mathbf{p}_\perp}{(2\pi)^3}
\psi(x,\mathbf{k}_\perp)\mathrm{Tr}[(-\slashed p_{\bar{q}}+m_{\bar{q}})\gamma^\mu\gamma^5(\slashed p_q+m_q)].
\label{eq:17}
\end{eqnarray}
The calculation of trace term is straightforward and the result for twist-2 DA is carried out
\begin{eqnarray}
\phi_2(x,\mu) = \frac{\sqrt{2N_c}}{f_M}\int_0^{\mu^2} \frac{d^2\mathbf{k}_\perp}{8\pi^3}\frac{\bar{x}m_q+xm_{\bar{q}}}{\sqrt{(\bar{x}m_q+xm_{\bar{q}})^2
+\mathbf{k}^2_\perp}}\psi(x,{\bf{k}}_{\perp}).
\label{eq:18}
\end{eqnarray}
Adopting the same procedure, the formula of twist-3 DAs are derived as
\begin{eqnarray}
\phi_{3,M}^P(x,\mu)&=&\frac{\sqrt{2N_c}}{f_M\mu_M}\int_0^{\mu^2} \frac{d^2\mathbf{k}_\perp}{16\pi^3}\frac{M_0^2-(m_q-m_{\bar{q}})^2}{\sqrt{(\bar{x}m_q+xm_{\bar{q}})^2
+\mathbf{k}^2_\perp}}\psi(x,{\bf{k}}_{\perp}),
\label{eq:19}
\end{eqnarray}
where the details are given in Appendix.
These DAs are usually expanded in terms of the Gegenbauer polynomials $C_n^{1/2}$ and $C_n^{3/2}$ as follows
\begin{eqnarray}
\phi_2(x,\mu)&=&\phi_{as}(x)\sum_0^\infty a_n(\mu)C_n^{3/2}(2x-1),\\
\phi_{3,M}^P(x,\mu)&=&\sum_0^\infty a_n^P(\mu)C_n^{1/2}(2x-1),
\label{eq:20}
\end{eqnarray}
where the asymptotic DA $\phi_{as}(x)=6x\bar{x}$ is well-known in the high energy limit. The important coefficients $a_n^{P}(\mu)$ are Gegenbauer moments which usually serve as essential input parameters in phenomenological researches and describe the discrepancy between DAs and the asymptotic one. The explicit formula for $a_n^{P}(\mu)$ is given by
\begin{eqnarray}
a_n^A(\mu)&=&\frac{4n+6}{3n^2+9n+6}\int_0^1 dx C_n^{3/2}(2x-1)\phi_2(x),\\
a_n^P(\mu)&=&(2n+1)\int_0^1 dx C_n^{1/2}(2x-1)\phi^P_3(x).
\label{eq:20p}
\end{eqnarray}
With the DAs of meson, we can obtain additional information about bound state's characteristics such as the $\xi$-moment
\begin{eqnarray}
\langle\xi^n\rangle = \int_0^1 dx\xi^n\phi(x),
\label{eq:22}
\end{eqnarray}
where $\xi$ represents the longitudinal discrepancy between constituent quarks in the bound state. Similarly, the nonperturbative quantity transverse moment is also obtained by
\begin{eqnarray}
\langle\mathbf{k}^n_\perp\rangle = \int_0^1 dx~\mathbf{k}^n_\perp~\phi(x),
\label{eq:21p}
\end{eqnarray}
which reflects the information about the transverse size of the bound state.
\section{Numerical Results and Discussion} \label{sec:3}

The model parameters for constituent quark masses and Gaussian parameter $\beta$ are listed in Table~\ref{tab:1}, which are mainly obtained by fitting the world averages of experimental data on mesonic decay constants from the Particle Data Group (PDG)~\cite{ParticleDataGroup:2022pth}. The explicit formula of decay constants in LFQM are displayed as
\begin{eqnarray}
f_P = \frac{\sqrt{2N_c}}{8\pi^3}\int_0^1 dx \int d^2\mathbf{k}_\perp\frac{xm_{\bar{q}}+\bar{x}m_q}{\sqrt{\mathbf{k}^2_\perp+(xm_{\bar{q}}+\bar{x}m_q)^2}}\psi(x,\mathbf{k}_\perp),
\label{eq:24}
\end{eqnarray}
for P-meson and
\begin{eqnarray}
f_V = \frac{\sqrt{2N_c}}{8\pi^3}\int_0^1 dx \int d^2\mathbf{k}_\perp\frac{(xm_{\bar{q}}+\bar{x}m_q)+\frac{2\mathbf{k}^2_\perp}{M_0+m_q+m_{\bar{q}}}}{\sqrt{\mathbf{k}^2_\perp+(xm_{\bar{q}}+\bar{x}m_q)^2}}
\psi(x,\mathbf{k}_\perp),
\label{eq:25}
\end{eqnarray}
for V-meson, respectively. Here are some comments about self-consistency of $f_V$ which involves zero-mode contribution and is not invariant by different ways of extraction. We have validated the effectiveness of replacement $M\to M_0$ in analysis of self-consistency problem in decay constants and weak transition form factors in our previous works~\cite{Chang:2019obq,Chang:2019mmh,Xu:2025aow}. The findings are unambiguous that the formally zero-mode contribution numerically vanish with the replacement $M\to M_0$ and various descriptions give the same numerical result. The fitting for parameters in this work are also performed with the replacement $M\to M_0$ and the parameters in the previous works are also present for comparison~\cite{Choi:2017uos}.

\begin{table*}[t]
  \caption{\label{tab:1}The inputs of Gaussian parameter $\beta$ [GeV] and quark masses where the scheme-I obtained by the combined fitting for mesonic decay constants~\cite{Xu:2025ntz} and the scheme-II obtained by the variational principle\cite{Choi:2009ai}.}
    \begin{tabular*}{\textwidth}{c@{\extracolsep{\fill}}ccccccccccc}
				\hline\hline
   This work &$m_q$ &
    $m_s$ & $m_c$ & $m_b$ & $\beta _{qq}$ &
  $\beta _{qs}$   & $\beta_{qc}$ & $\beta_{cs}$ & $\beta _{qb}$ & $\beta _{bs}$ &
  $\beta _{bc}$ \\
    \hline
    I~\cite{Xu:2025ntz} &0.25 &
    0.50 & 1.80 & 5.10 & 0.321 & 0.352 & 0.465 &
    0.522 & 0.535 & 0.594 & 0.883\\
    \hline
       II~\cite{Choi:2009ai}  &0.22 &
    0.45 & 1.80 & 5.20 & 0.367 & 0.388 & 0.468 &
    0.501 & 0.526 & 0.571 & 0.807\\
		\hline
    \end{tabular*}
\end{table*}

Before the numerical analysis, it is necessary to clarify normalization for DAs in Eqs.~\ref{eq:18}-\ref{eq:19} which should satisfy
\begin{eqnarray}
\int_0^1 dx~\phi_2(x)=\int_0^1 dx~\phi^P_3(x)=1.
\label{eq:26}
\end{eqnarray}
It is found that the normalization condition for twist-2 DA $\phi_2(x)$ is strictly established, whereas the one for twist-3 DA $\phi^P_3(x)$ is not rigorously satisfied. The issue seemingly originates from the factor $\mu_M=M^2/(m_q+m_{\bar{q}})$ in the formula, which should be treated as $M_0^2/(m_q+m_{\bar{q}})$ with the replacement $M\to M_0$. Then one can check the DAs are self-normalized. Accordingly, the numerical results below are carried out under the replacement $M\to M_0$.

According to QCD, the scale $\mu$ separates perturbative and nonperturbative regimes and a cutoff via $| \mathbf{k}_{\perp}|\leq\mu$ in transverse momentum integration is the usual way to carry out numerical results in the LFQM. In several previous literature, a value of $\mu^2\sim1$~GeV$^2$ is commonly adopted in the studies for pion and kaon DAs, which is also chosen as the renormalization scale. It should be noted that $\mu^2\sim10$~GeV$^2$ for heavy mesons.
\subsection{The Pions and Kaons}\label{sec:pika}
\begin{figure*}[htb]
\centering
\includegraphics[width=8.1cm,height=7cm]{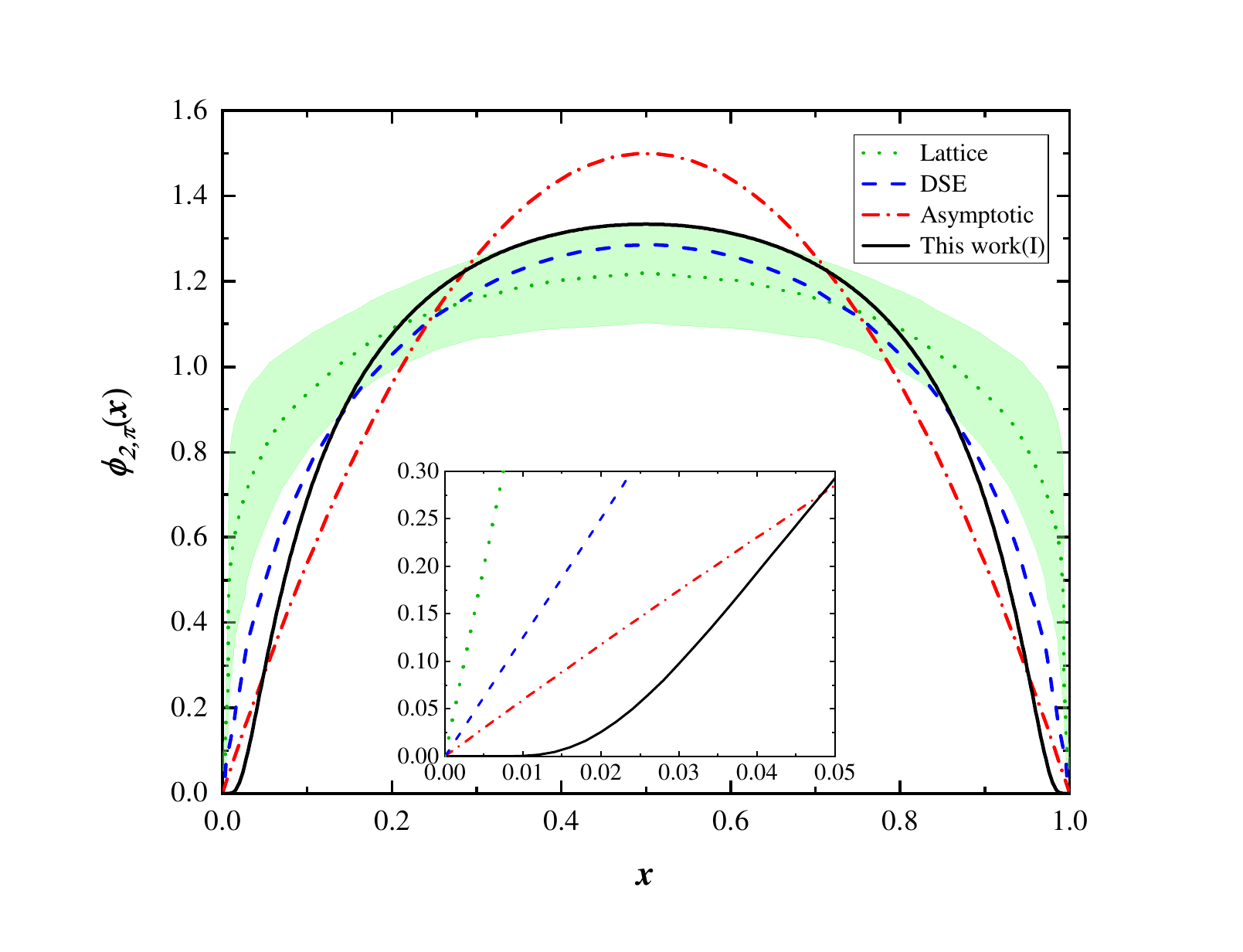}
\includegraphics[width=8.1cm,height=7cm]{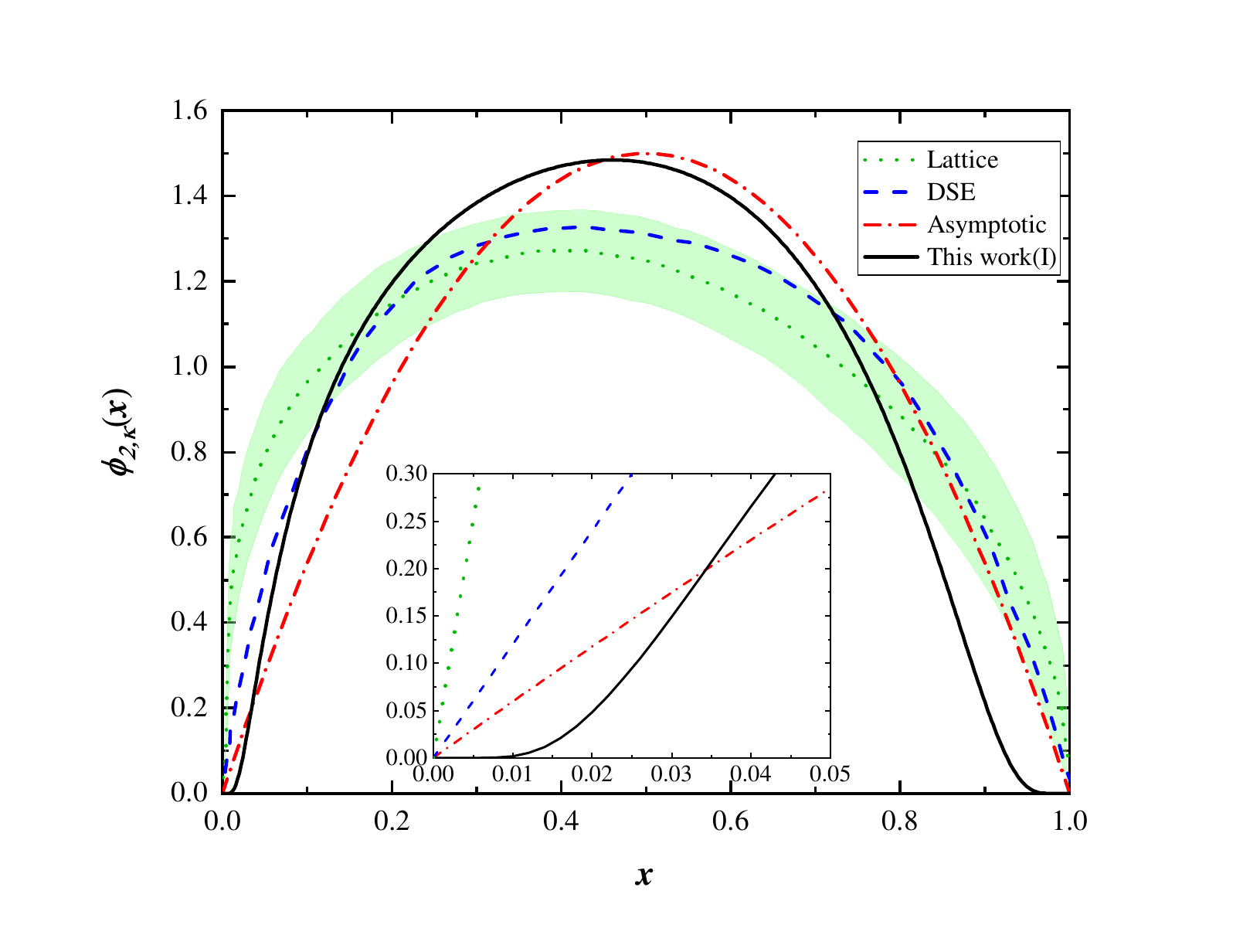}
\includegraphics[width=8.1cm,height=7cm]{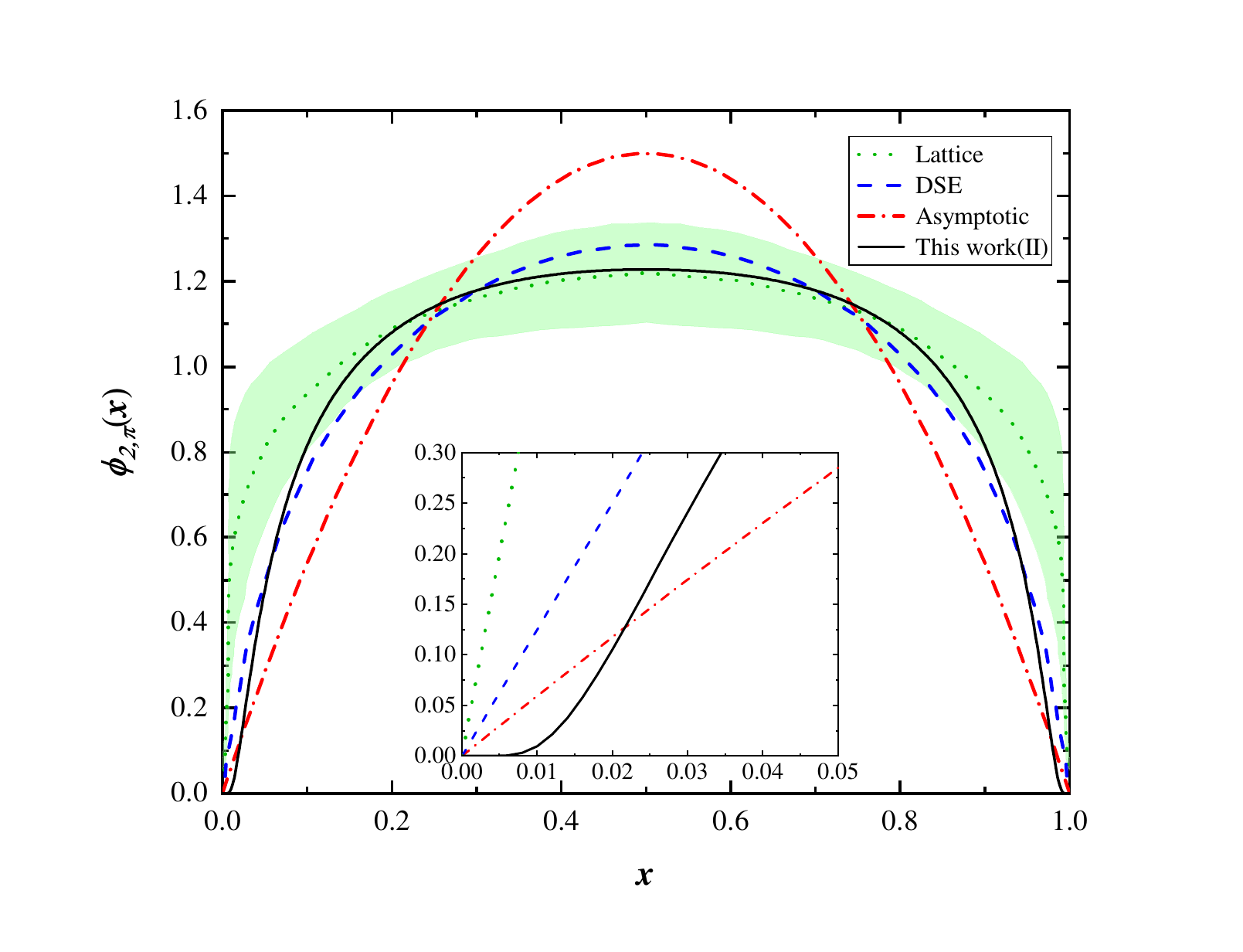}
\includegraphics[width=8.1cm,height=7cm]{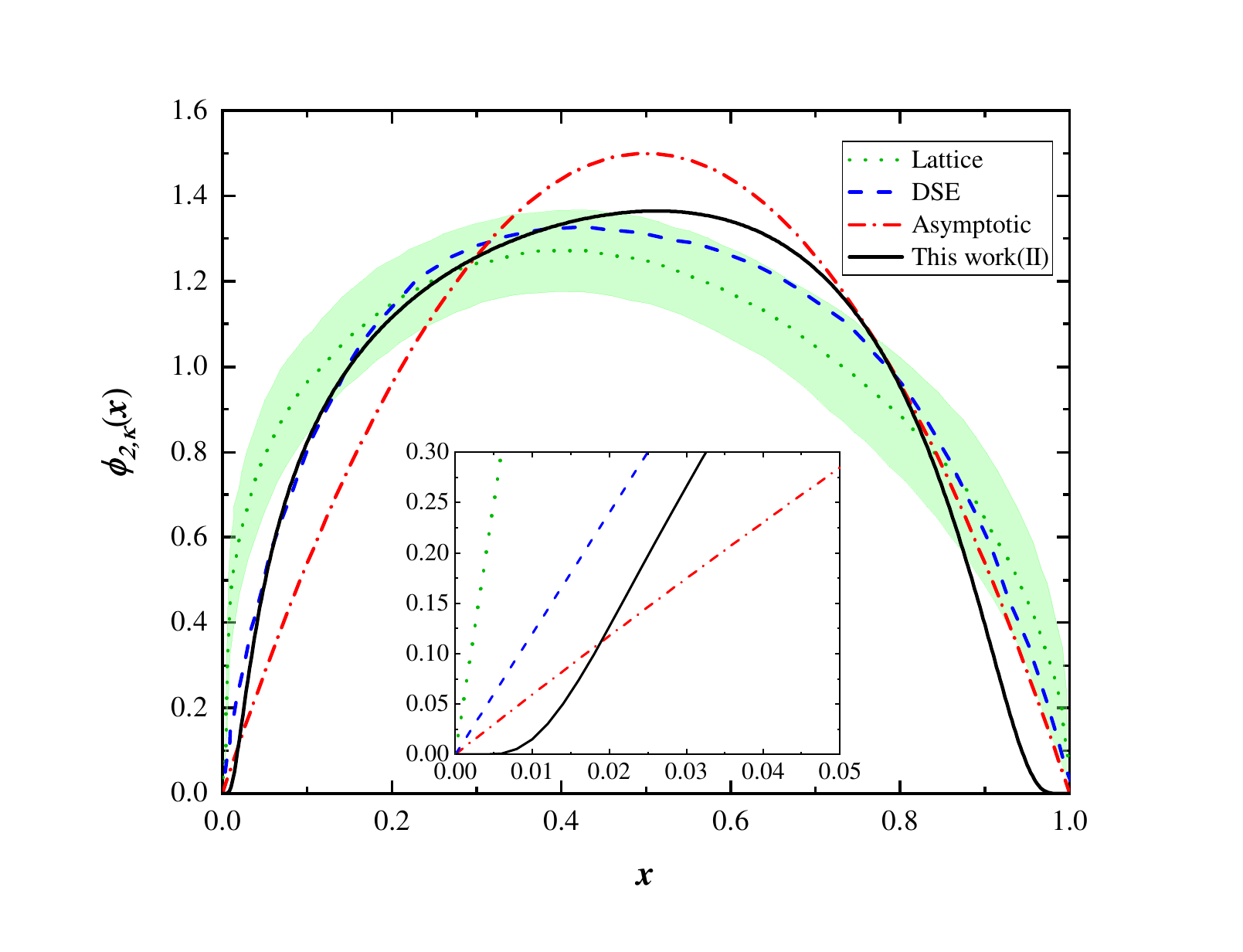}
\caption{The twist-2 DAs of the pion and kaon obtained from the LFQM under scheme-I and II, compared with lattice data from Ref.~\cite{LatticeParton:2022zqc} and DSE from Ref.~\cite{Roberts:2021nhw}.}
\label{1}
\end{figure*}

\begin{table*}[!]
  \caption{\label{tab:2} The first four Gegenbauer moments of twist-2 DAs for pion and kaon at $\mu^2 = 1~ \text{GeV}^2$.}
	\renewcommand{\tabcolsep}{0.2pc}
    \begin{tabular*}{\textwidth}{l @{\extracolsep{\fill}}cccccc}
				\hline\hline
       Model & $a_{2,\pi}$ & $a_{4,\pi}$ & $a_{1,K}$ & $a_{2,K}$ & $a_{3,K}$ & $a_{4,K}$ \\[4pt]
    \hline
     This work(I) & $0.054^{+24}_{-28}$ & $-0.034^{+10}_{-07}$ & $-0.138^{+27}_{-31}$ &$ -0.023^{+10}_{-04}$  & $-0.041^{+03}_{-01}$ & $-0.032^{+05}_{-02}$ \\
     This work(II) & 0.126 & -0.004 & -0.089 & 0.032   & -0.057 & -0.026 \\
     Lattice~\cite{LatticeParton:2022zqc} & $0.258^{+70}_{-52}$ & $0.122^{+46}_{-31}$ & $-0.108^{+14}_{-51}$ & $0.170^{+14}_{-44}$ & $-0.043^{+06}_{-22}$ & $0.073^{+08}_{-21}$ \\
     Lattice~\cite{Chen:2023byr,RQCD:2019osh} & $0.116^{+19}_{-20}$ & -- & $0.0525^{+31}_{-33}$ & $0.106^{+15}_{-16}$ & -- & -- \\
     NL$\chi$QM~\cite{Nam:2006sx}& 0.053 & -0.061&-0.010&-0.037&0.007&-0.038 \\
     NL$\chi$QM~\cite{Nam:2006au}& 0.029 & -0.046& 0.096&-0.051&-0.008&-0.040 \\
     DSE-DB~\cite{Chang:2013pq,Raya:2015gva} & 0.149 & 0.076&--&--&--&-- \\
     DSE-RL~\cite{Chang:2013pq,Raya:2015gva} & 0.233 & 0.112&--&--&--&-- \\
     SVZSR~\cite{Zhong:2014jla} &$0.403^{+93}_{-93}$ & $0.320^{+83}_{-83}$& -- & -- &  -- & --\\
     QCDSR~\cite{Khodjamirian:2004ga} &$0.26^{+21}_{-09}$ &--& $0.05^{+02}_{-02}$ & $0.27^{+37}_{-12}$ &  -- & --\\
     QCDSR~\cite{Ball:2006wn} &$0.25^{+15}_{-15}$ & --& $0.06^{+03}_{-03}$ & $0.30^{+15}_{-15}$ &  -- & --\\
     LCSR~\cite{Agaev:2010aq,Agaev:2012tm} &0.10-0.14 & 0.10-0.18& -- & -- &  -- & --\\
     LCSR~\cite{Cheng:2020vwr} &0.22-0.33 & 0.12-0.25& -- & -- &  -- & --\\
     [1pt]
		 \hline
    \end{tabular*}
\end{table*}

\begin{figure*}[htb]
\centering
\includegraphics[width=8.1cm,height=7cm]{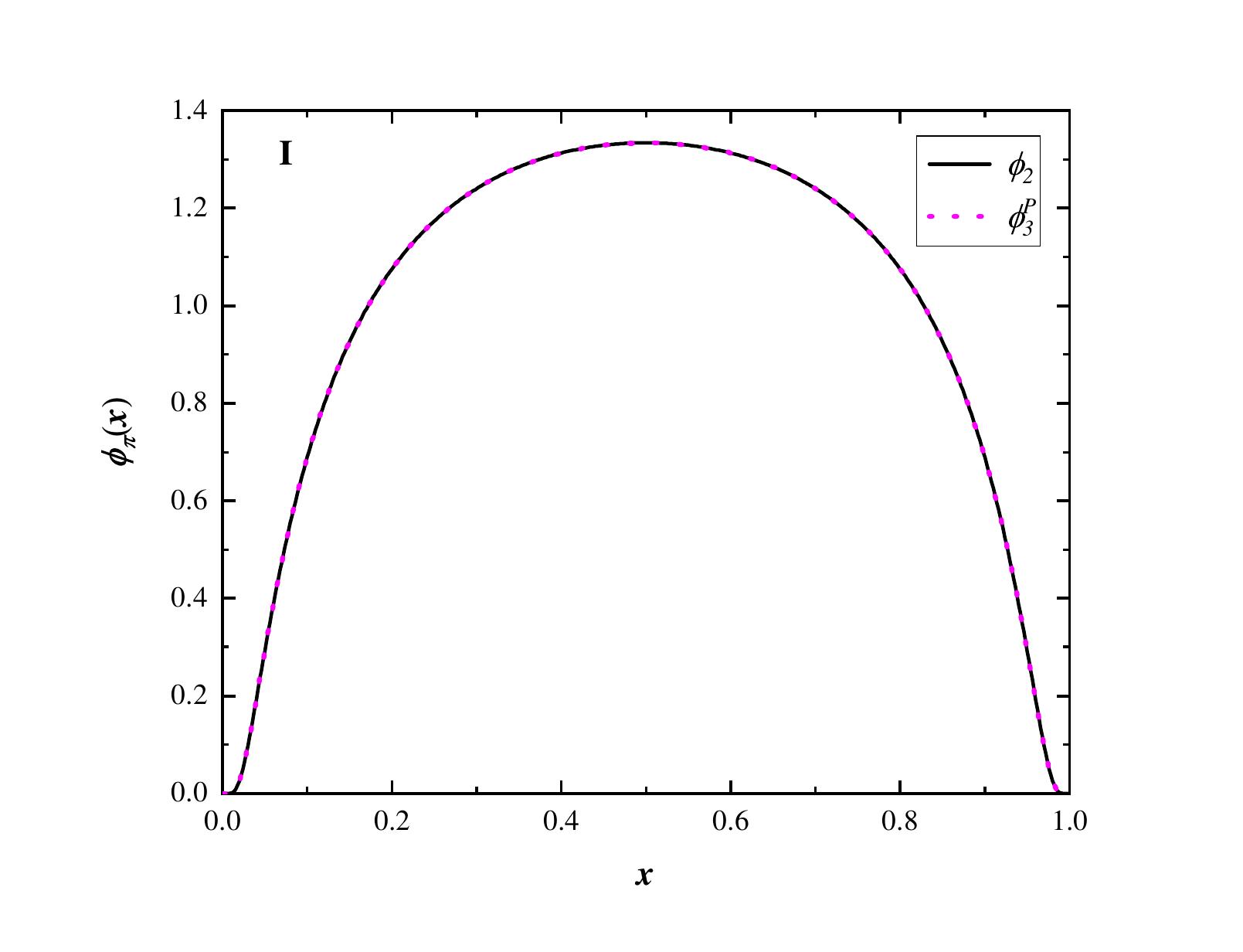}
\includegraphics[width=8.1cm,height=7cm]{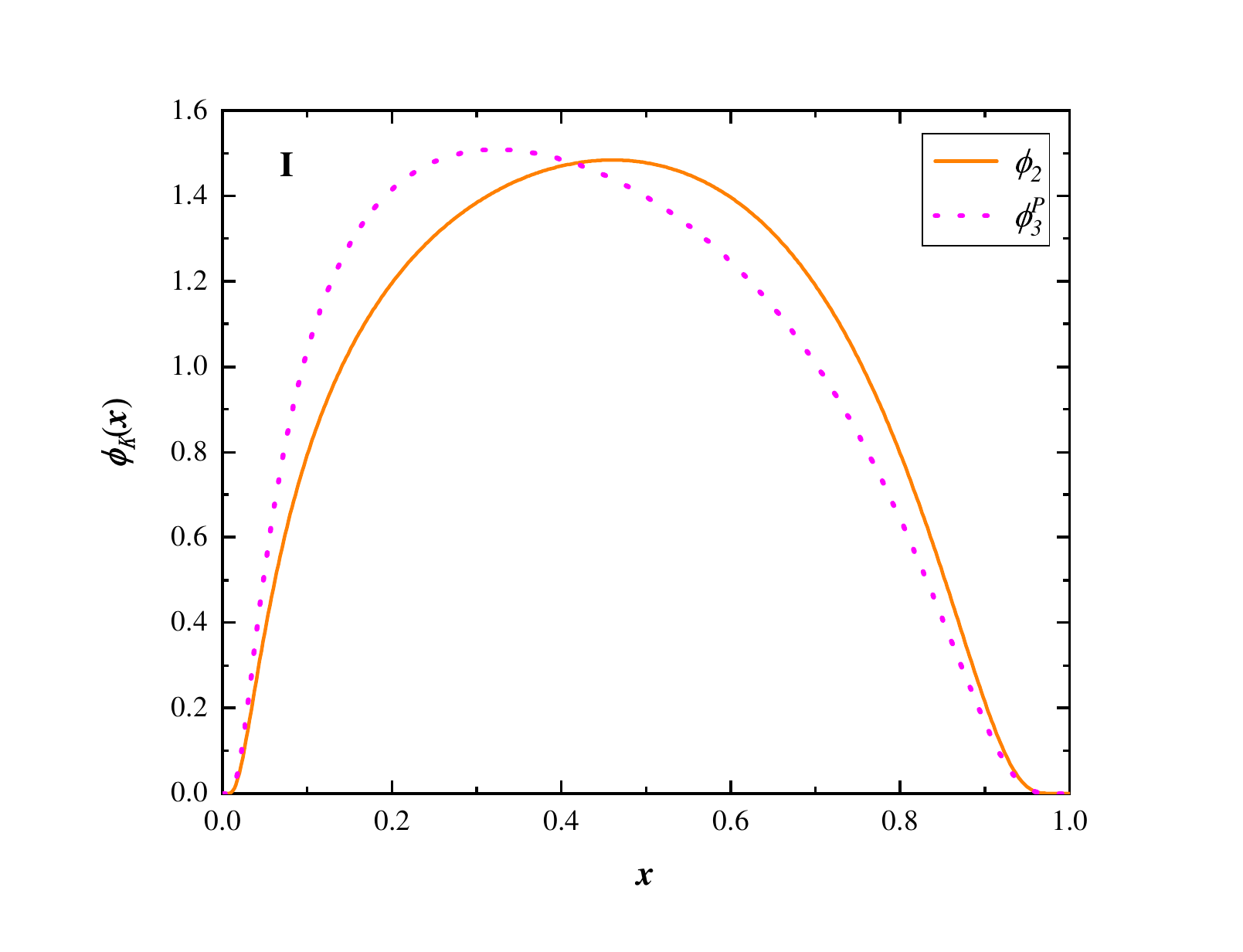}
\includegraphics[width=8.1cm,height=7cm]{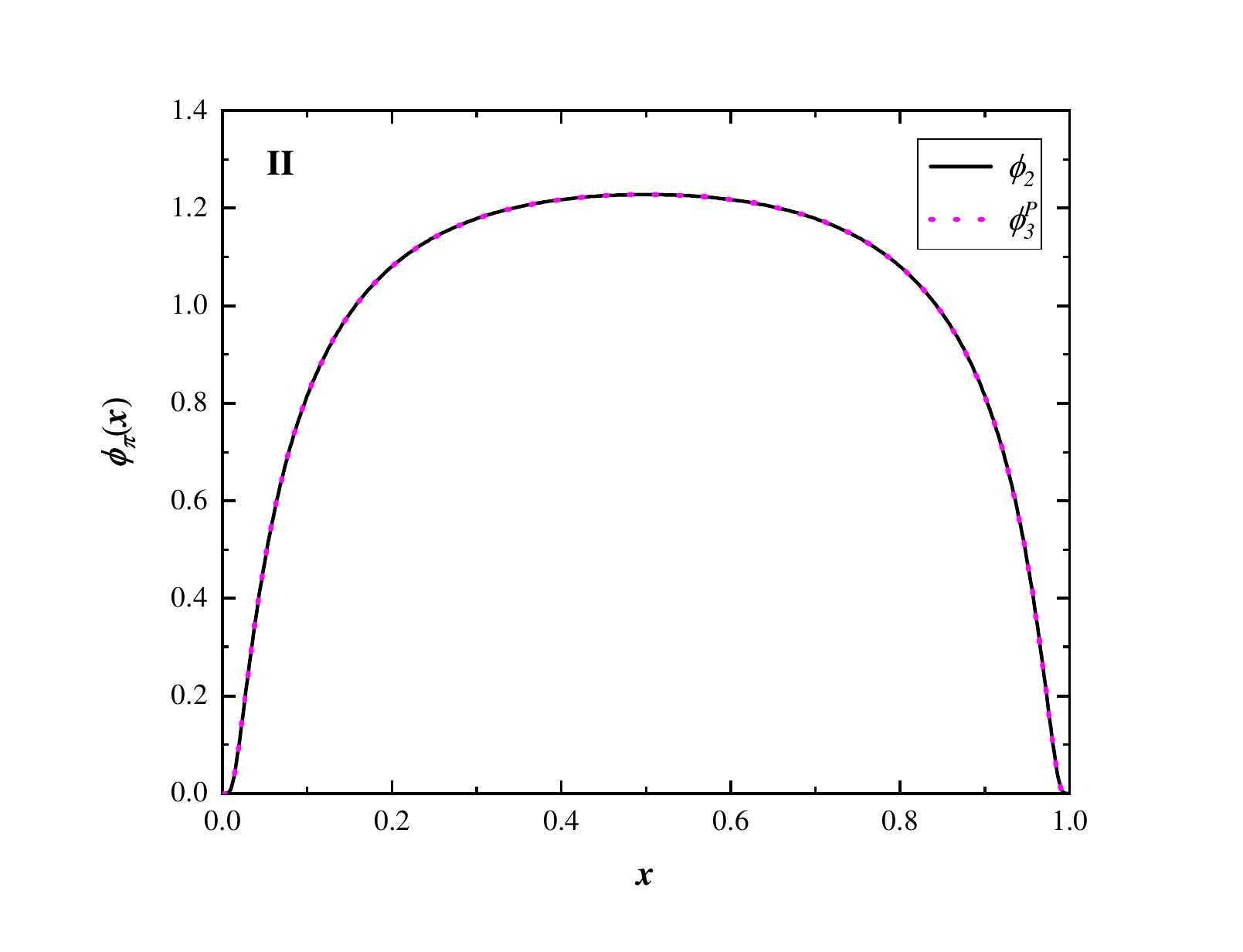}
\includegraphics[width=8.1cm,height=7cm]{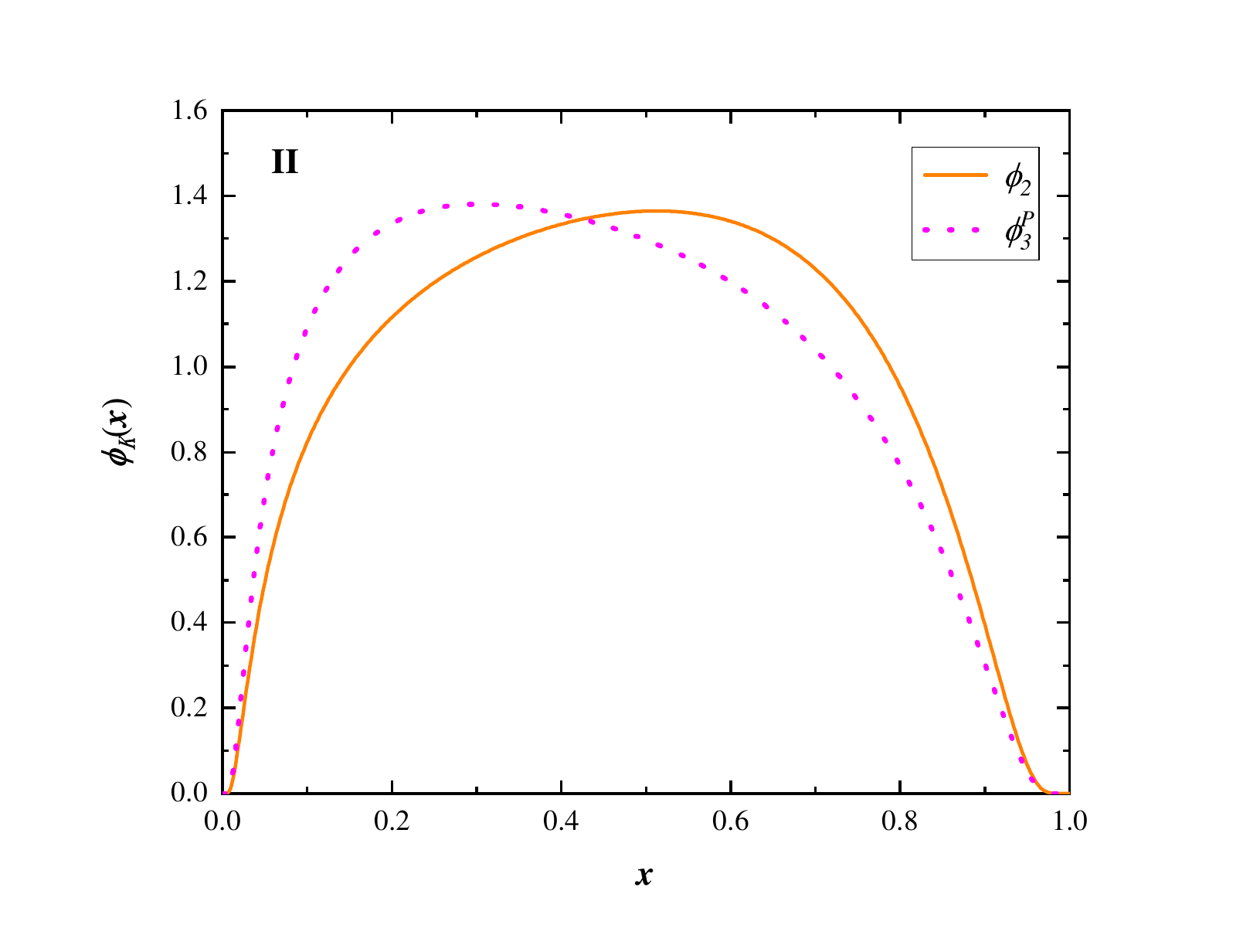}
\caption{The twist-2 and twist-3 DAs of the pion and kaon obtained from the LFQM under scheme-I and II.}
\label{2}
\end{figure*}

\begin{table*}[!]
  \caption{\label{tab:3}The first four Gegenbauer moments of twist-3 DAs for pion and kaon at $\mu^2 = 1~ \text{GeV}^2$.}

	\renewcommand{\tabcolsep}{0.5pc}
    \begin{tabular*}{\textwidth}{l @{\extracolsep{\fill}}cccccc}
				\hline\hline
       Model & $a_{2,\pi}^P$ & $a_{4,\pi}^P$ & $a_{1,K}^P$ & $a_{2,K}^P$ & $a_{3,K}^P$ & $a_{4,K}^P$ \\[4pt]
    \hline
     This work(I) & -0.859 & -0.273 & -0.435 & -0.939  & 0.304 & -0.169 \\
     This work(II) & -0.669 & -0.340 & -0.377 & -0.791   & 0.159 & -0.253 \\
     QCDSR~\cite{Ball:2006wn} &  0.437 & -0.071 &  0.184 &  0.271   & 0.395 & -0.247 \\
     NL$\chi$QM~\cite{Nam:2006mb}& -0.431 & -0.556& 0.024&-0.647&-0.037&-0.372 \\[1pt]
		 \hline
    \end{tabular*}
\end{table*}

The results of twist-2 DA $\phi_2(x)$ for pion and kaon is shown in Fig.~\ref{1}, where the asymptotic form and lattice data from Ref.~\cite{LatticeParton:2022zqc} and DSE data from Ref.~\cite{Roberts:2021nhw} are also presented for comparison. Some comments are provided as follows:
\begin{itemize}
\item As anticipated by SU(2) symmetry, the DAs of pion exhibits perfect symmetry about $x=0.5$. There are significant deviations between the results from LFQM and that of asymptotic form or lattice data: The shape of $\phi_{2,\pi}(x)$ from LFQM is slight broader than the asymptotic one but much narrower than the lattice one, which may be explained by the dynamic chiral symmetry breaking effect~\cite{LatticeParton:2022zqc,Chang:2013pq}. It is noticeable that the area under the lattice curve appears larger than that of LFQM and asymptotic form, where the latter are exactly equal to unity, i.e., the normalization condition for the former may not be satisfied. Moreover, the endpoint behaviors are distinctly different as shown in Fig.~\ref{1}. The endpoint contribution in the DA from lattice is the most significant, which may stem from the ineffectiveness of the lattice approach near the endpoints and relates to phenomenologically implemented extrapolation as a consequence. The suppression effect in our results is the strongest and similar to the Sudakov suppression of the soft contribution~\cite{Choi:2007yu,Li:1992nu,Li:1992ce}, which appears more physically plausible for hard exclusive processes than that of asymptotic and other forms. The distinctive performance of LFQM's DAs is an inherent feature of the exponential Gaussian-type wave function, which will further affect the asymptotic behavior of electromagnetic form factors (EMFFs), the details see Ref.~\cite{Xu:2025ntz}.

\item As shown in the right panel of Fig.~\ref{1}, which displays the details of the kaon twist-2 DA $\phi_{2,K}(x)$. the curves from different theoretical approaches exhibit significantly different shapes. Owing to SU(3) flavor symmetry breaking, all descriptions show appreciable asymmetry, with the lattice result being the most pronounced. This indicates that the lighter constituent quark carries a smaller fraction of the longitudinal momentum than the heavier one, and the $\xi$-moment is expected to be nonzero. The description of LFQM implies a higher probability at the most probable momentum distribution and a lower probability at near endpoint.

\item The twist-3 DAs of pion and kaon are shown in Fig.~\ref{2}, along with the corresponding twist-2 DAs for comparison. With the replacement $M\to M_0$, the shapes of the curves for $\phi_{2,\pi}(x)$ and $\phi^P_{3,\pi}(x)$ are the same as each other. The SU(3) flavor symmetry breaking effect is significantly stronger in $\phi^P_{3,K}(x)$ than in $\phi_{2,K}(x)$.

\end{itemize}

For the twist-2 DA, the curve under scheme-II exhibits a flatter profile compared to scheme-I, with a significantly reduced peak value. This morphological difference directly leads to the pronounced discrepancy in the Gegenbauer moments obtained under the two schemes.
Before proceeding to specific numerical analysis, we first provide a brief commentary on the current research status. As shown in Table~\ref{tab:2}, despite intensive studies of the pion have been performed, there are significant controversies in the existing results. Predictions vary widely across different theoretical models, and even within the same model, distinct values are obtained under different implementation schemes. The research on kaon and heavier mesons is relatively scarce, and predictions from different theories also show considerable divergence.

For the Gegenbauer moment $a_{2,\pi}$, the scheme-I result is less than half of that obtained under scheme-II, in accordance with the smaller deviation of the DA from its asymptotic form under scheme-I, as illustrated in Fig.~\ref{1}. In addition, guided by the fitting ranges of the Gaussian parameter $\beta$ and the constituent quark masses established in our previous works~\cite{Xu:2025ntz}, we estimate the corresponding theoretical uncertainties for the scheme-I Gegenbauer moments in Table~\ref{tab:2}. The same method is employed to assess the uncertainties quoted in Tables~\ref{tab:4} and~\ref{tab:5}. The $a_{2,\pi}$ under scheme-I well consists with predictions of the platykurtic~(0.056, 0.115)~\cite{Stefanis:2020rnd} and NL$\chi$QM, and also falls within the range forecasted by QCDSR. However, this result is significantly lower than lattice results from recent works~\cite{Chen:2023byr,LatticeParton:2022zqc,RQCD:2019osh}. In contrast, the value of $a_{2,\pi}$ obtained under scheme-II is in good agreement with both lattice QCD results~\cite{RQCD:2019osh} and DSE calculations~\cite{Chang:2013pq}, and also lies within the broad range suggested by data-driven LCSR analyses~\cite{Agaev:2010aq,Agaev:2012tm}.
For the Gegenbauer moments $a_{4,\pi}$, both schemes yield negative values and the results exhibit higher-order suppression compared to the $a_{2,\pi}$. The $a_{4,\pi}$ under scheme-I aligns well with predictions from the platykurtic~(-0.049, 0.015)\cite{Stefanis:2020rnd} and the NL$\chi$QM, while show significant discrepancies with QCDSR and lattice calculations. As shown in Table~\ref{tab:3}, the values of Gegenbauer moments of twist-3 DA under both two schemes are substantially larger than those of twist-2 DA and are predominantly negative.
\begin{table*}[!]
  \caption{\label{tab:4}The first four $\xi$-moments of twist-2 and twist-3 DAs for the pion and kaon.}

	\renewcommand{\tabcolsep}{0.15pc}
    \begin{tabular*}{\textwidth}{l @{\extracolsep{\fill}}cccccc}
				\hline\hline
       Model & $\langle\xi^2\rangle_\pi$ & $\langle\xi^4\rangle_\pi$  & $\langle\xi^1\rangle_K$  & $\langle\xi^2\rangle_K$ & $\langle\xi^3\rangle_K$ & $\langle\xi^4\rangle_K$ \\[4pt]
    \hline
     This work(I) & $0.218^{+09}_{-09}$ & $0.094^{+05}_{-07}$ &$-0.083^{+16}_{-18}$ &$0.191^{+04}_{-04}$ & $-0.043^{+07}_{-07}$ & $0.076^{+03}_{-02}$ \\
     This work(II) & 0.241 & 0.113 & -0.054 & 0.208   & -0.034 & 0.089 \\
     QCDSR~\cite{Guo:1991eb} &  0.360 & 0.170 &  0.150 &   0.340   & 0.075 & 0.300 \\
     NL$\chi$QM~\cite{Nam:2006au} &  0.210 &  0.090 &  0.057 &  0.182   & 0.023 &  0.070 \\\hline\hline
      Model & $\langle\xi^2\rangle_\pi^P$ & $\langle\xi^4\rangle_\pi^P$  & $\langle\xi^1\rangle_K^P$  & $\langle\xi^2\rangle_K^P$ & $\langle\xi^3\rangle_K^P$ & $\langle\xi^4\rangle_K^P$ \\[4pt]
    \hline
     This work(I) & 0.218 & 0.094 &-0.145 &0.205 & -0.070 & 0.087 \\
     This work(II) & 0.241 & 0.113 & -0.125 & 0.223   & -0.066 & 0.100 \\
     QCDSR~\cite{Ball:2006wn} &  0.386 & 0.245 &  0.386 &   0.245   & 0.395 & -0.247 \\
     NL$\chi$QM~\cite{Nam:2006mb}&  0.276 &  0.137& 0.023&-0.647&-0.037&-0.372 \\[1pt]
		 \hline
    \end{tabular*}
\end{table*}

\begin{table}[!]
  \caption{\label{tab:5}The $2$-th transverse moment $\sqrt{\langle \mathbf{k}^2_\perp\rangle}$ of twist-2 and twist-3 DAs for the pion and kaon.}
	\renewcommand{\tabcolsep}{1pc}
    \begin{tabular}{lcccc}
				\hline\hline
       Model & $\sqrt{\langle \mathbf{k}^2_\perp\rangle_{\phi_{2,\pi}}}$ &  $\sqrt{\langle \mathbf{k}^2_\perp\rangle_{\phi_{2,K}}}$  &  $\sqrt{\langle \mathbf{k}^2_\perp\rangle_{\phi_{3,\pi}^P}}$  & $\sqrt{\langle \mathbf{k}^2_\perp\rangle_{\phi_{3,K}^P}}$  \\[4pt]
    \hline
     This work(I) & $0.372^{+15}_{-15}$ & $0.424^{+17}_{-17}$ &$0.372^{+15}_{-15}$ &$0.424^{+17}_{-17}$  \\
     This work(II) & 0.412 & 0.457 & 0.412 & 0.457 \\
	pQCD~\cite{Chai:2025xuz} &  $0.36^{+02}_{-02}$ &  $0.55^{+07}_{-07}$ &  $0.40^{+02}_{-02}$  &  $0.53^{+07}_{-07}$ \\[1pt] \hline
    \end{tabular}
\end{table}

The numerical results of $\xi$- and $\mathbf{k}_{\perp}$-moments are listed in Tables~\ref{tab:4}-\ref{tab:5}, respectively. For the pion DAs, due to the SU(2) symmetry, the odd $\xi$-moments are zero. In addition, our results show the even $\xi$-moments are exactly the same for twist-2 and twist-3, which is reasonable given that $\phi_{2,\pi}=\phi^P_{3,\pi}$ under the $M\to M_0$ replacement. In case of kaon DAs, the odd $\xi$-moments are nonzero and the values of twist-3 exhibit generally greater magnitudes compared to that of twist-2. This indicates the SU(3) flavor breaking effect in the twist-3 DA is significantly larger than the one in the twist-2 DA, which confirms the finding above. Unexpectedly, Table~\ref{tab:5} reveals that results of 2-th transverse moment $\langle \mathbf{k}^2_{\perp}\rangle$ are identical for twist-2 and -3 DAs, which hold in both cases of pion and kaon. Contrary to previous assessments, this unique finding presents a novel perspective.

\subsection{The heavy flavor mesons}
\begin{figure*}[htb]
\centering
\includegraphics[width=8.1cm,height=7cm]{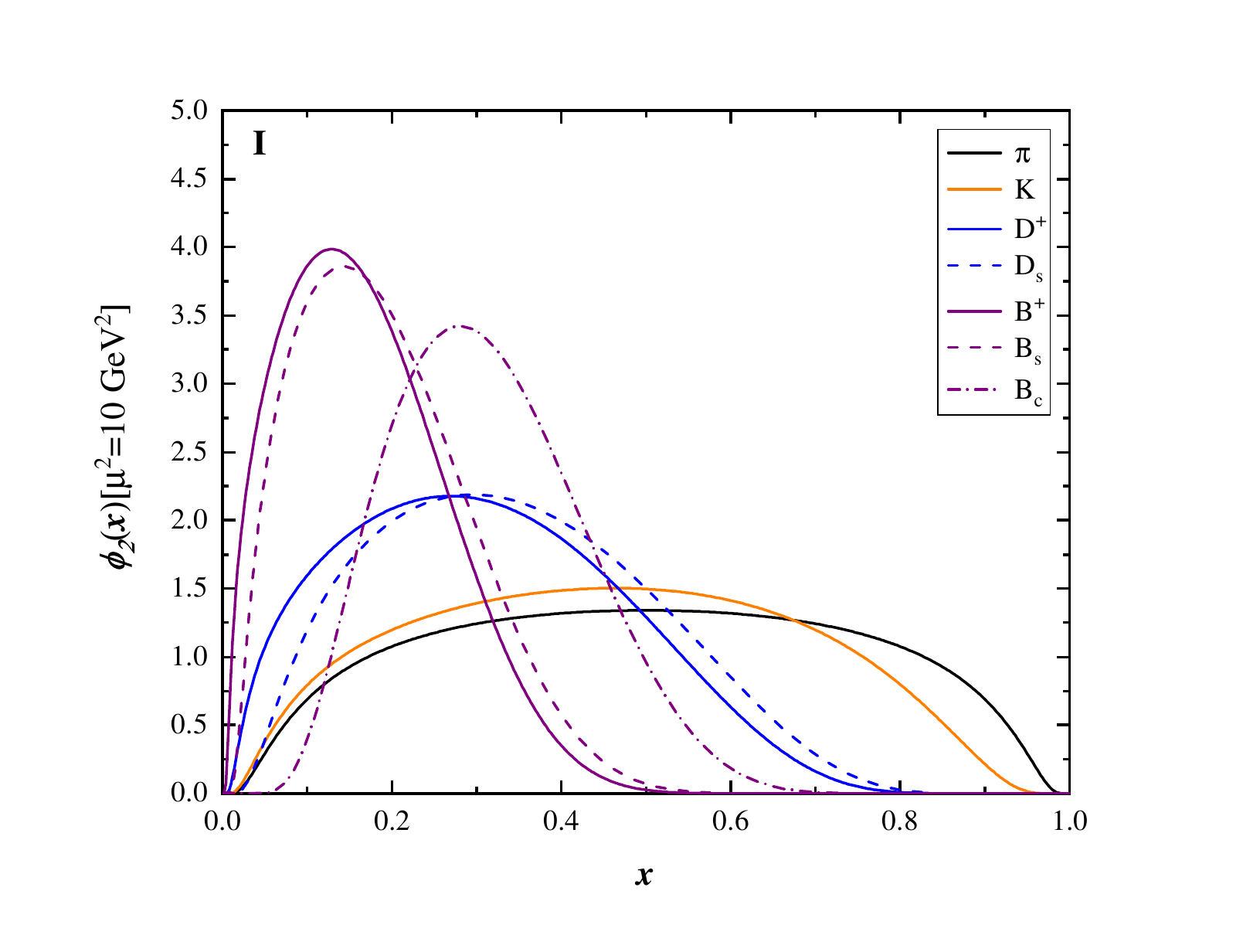}
\includegraphics[width=8.1cm,height=7cm]{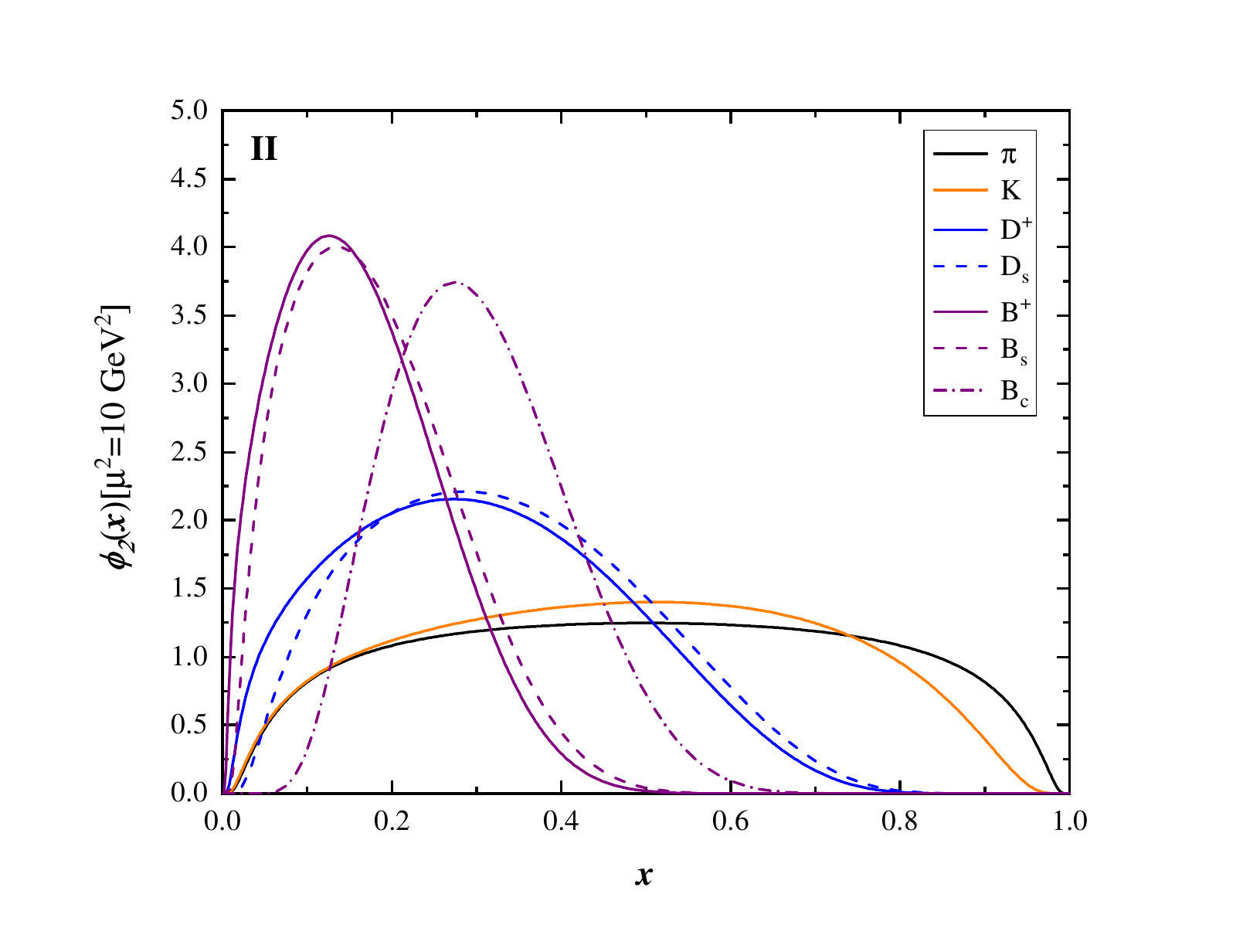}
\caption{The twist-2 DAs of the P-mesons obtained from the LFQM under scheme-I and II.}
\label{3}
\end{figure*}

\begin{figure*}[htb]
\centering
\includegraphics[width=8.1cm,height=7cm]{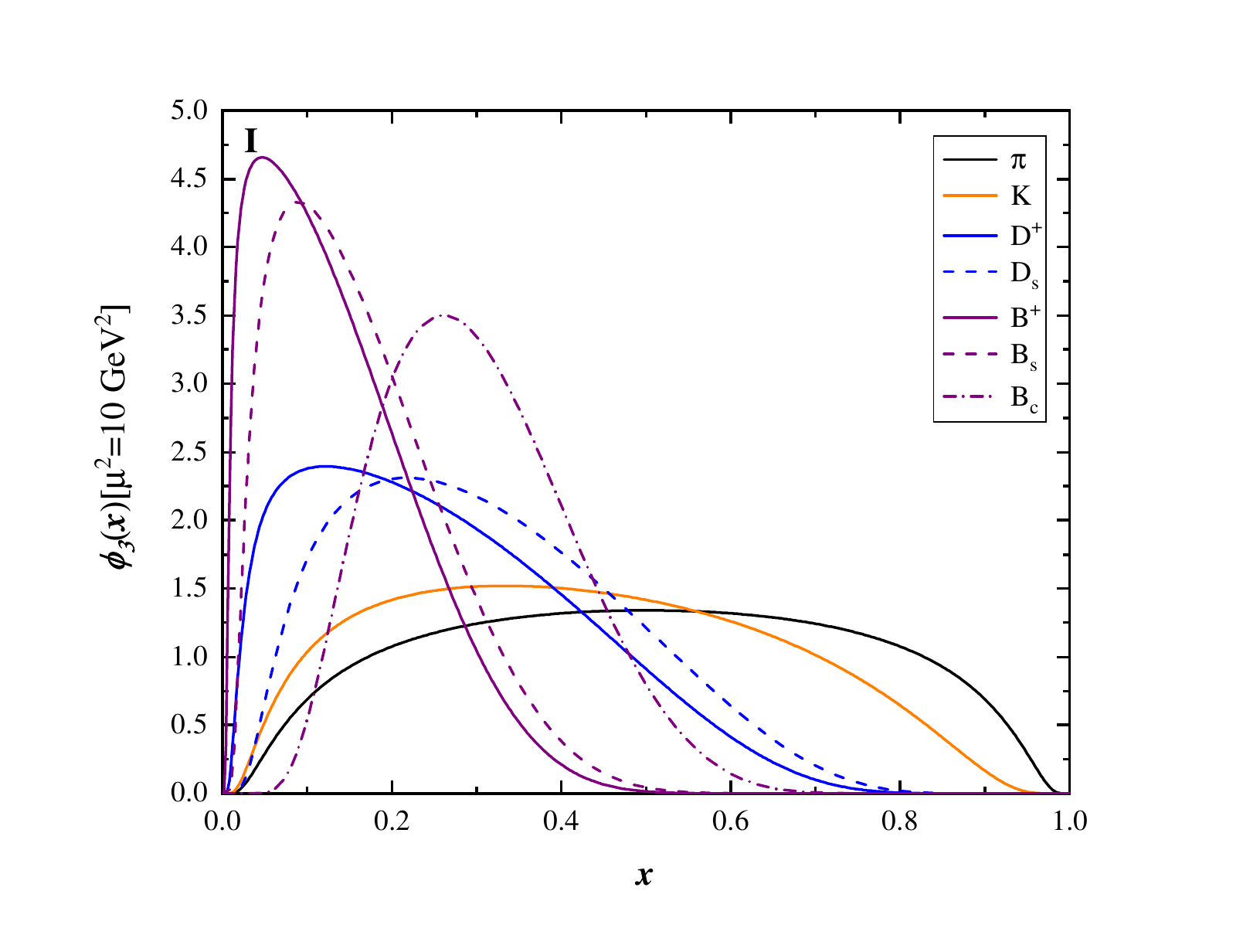}
\includegraphics[width=8.1cm,height=7cm]{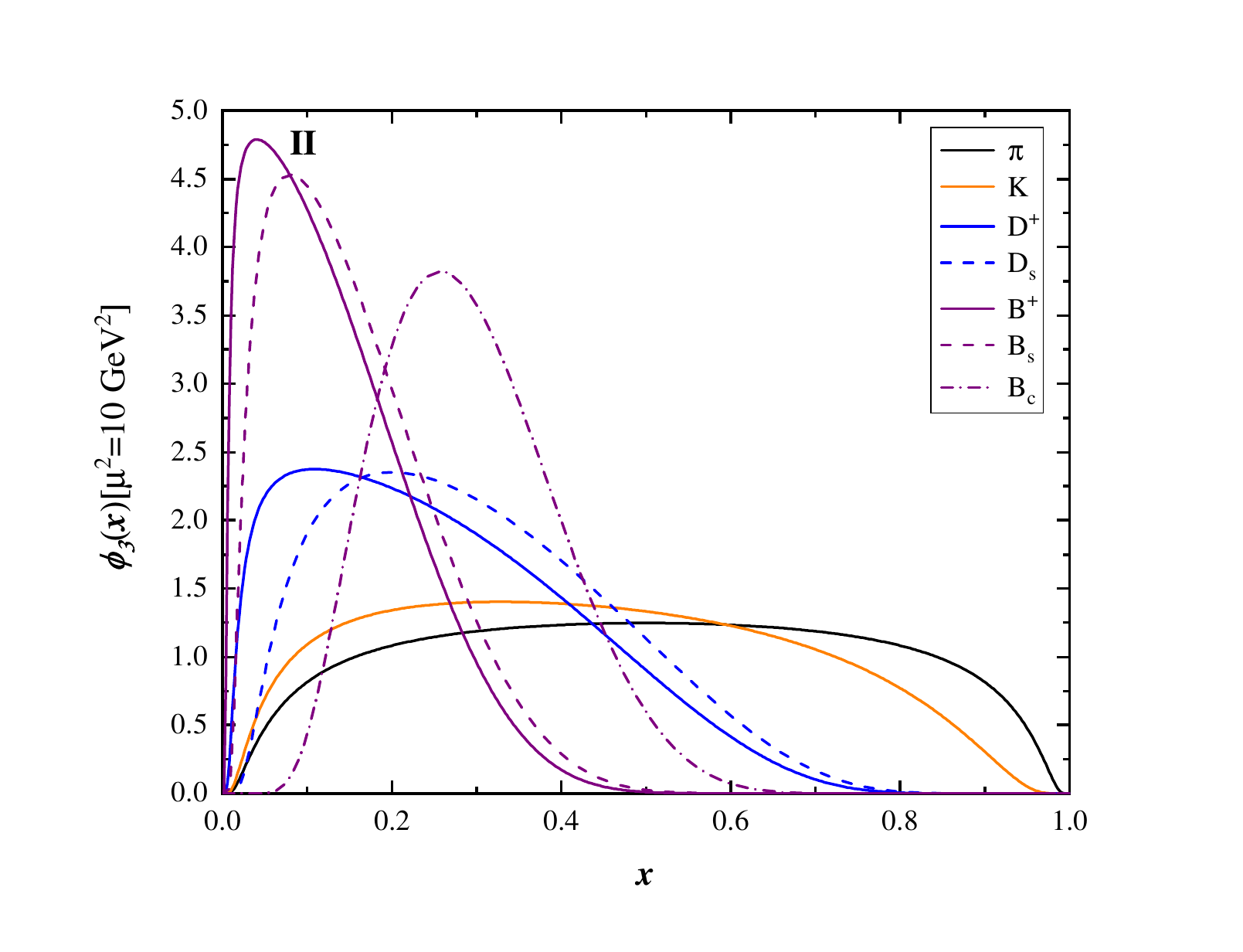}
\caption{The twist-3 DAs of the P-mesons obtained from the LFQM under scheme-I and II.}
\label{4}
\end{figure*}

\begin{table}[h!]
  \caption{\label{tab:6}The first four Gegenbauer moments of twist-2 DAs for P-mesons at $\mu^2 = 10~ \text{GeV}^2$.}
\renewcommand{\tabcolsep}{1.5pc}
    \begin{tabular}{cccccc}\hline\hline
    Mesons & ~~I/II &~~$a_1$&~~$a_2$ &~~ $a_3$ & ~~$a_4$ \\[4pt]\hline
    \multirow{2}{*}{$D^+$}&I & -0.628 &~~0.129 &~~ 0.010 &~~ 0.023 \\
    &II & -0.626 & ~~0.132 & ~~0.001 & ~~~0.034 \\
     \cline{1-6}
    \multirow{2}{*}{$D_s$}&I & -0.517 &~~0.004 &~~ 0.090 &~~ -0.030 \\
    &II & -0.549 & ~~0.024 & ~~~0.081 &~~ 0.028 \\
     \cline{1-6}
    \multirow{2}{*}{$B^+$}&I & -1.100 &~~0.791 &~~ -0.376 &~~ 0.101 \\
    &II & -1.116 & ~~0.823 &~~ -0.411 &~~ 0.124 \\
     \cline{1-6}
    \multirow{2}{*}{$B_s$}&I & -1.030 &~~0.646 &~~ -0.207 &~~ -0.039 \\
    &II & -1.060 & ~~0.709 &~~ -0.271 & ~~-0.001 \\
     \cline{1-6}
    \multirow{2}{*}{$B_c$}&I & -0.622 &~~0.035 &~~ 0.302 &~~ -0.165 \\
    &II & -0.656 & ~~-0.011 &~~ 0.326 &~~ -0.203 \\
     \cline{1-6}
     \hline
    \end{tabular}

\end{table}

\begin{table}[!]
  \caption{\label{tab:7}The first four Gegenbauer moments of twist-3 DAs for P-mesons at $\mu^2 = 10~ \text{GeV}^2$.}
\renewcommand{\tabcolsep}{1.5pc}
    \begin{tabular}{cccccc}\hline\hline
    Mesons & ~~I/II &~~$a_1^P$&~~$a_2^P$ &~~ $a_3^P$ & ~~$a_4^P$ \\[4pt]\hline
    \multirow{2}{*}{$D^+$}&I & -1.410 &~~-0.074 &~~ 0.914 &~~ -0.534 \\
    &II & -1.431 & ~~-0.016 & ~~0.830 & ~~~-0.466 \\
     \cline{1-6}
    \multirow{2}{*}{$D_s$}&I & -1.121 &~~-0.697 &~~ 1.280 &~~ -0.480 \\
    &II & -1.190 & ~~-0.577 & ~~~1.266 &~~ -0.558 \\
     \cline{1-6}
    \multirow{2}{*}{$B^+$}&I &-2.164 &~~1.662 &~~ -0.207 &~~ -0.888 \\
    &II & -2.195 & ~~1.762 &~~ -0.351 &~~ -0.773 \\
     \cline{1-6}
    \multirow{2}{*}{$B_s$}&I & -2.008 &~~1.136 &~~ 0.577 &~~ -1.574 \\
    &II & -2.066 & ~~1.314 &~~ 0.360 & ~~-1.469 \\
     \cline{1-6}
    \multirow{2}{*}{$B_c$}&I & -1.198 &~~-0.945 &~~ 2.159 &~~ -0.855 \\
    &II & -1.249 & ~~-0.894 &~~ 2.284 &~~ -1.052 \\
     \cline{1-6}
     \hline
    \end{tabular}
\end{table}
\begin{table}[!]
  \caption{\label{tab:8}The first four $\xi$-moment $\langle\xi^n\rangle$ of twist-2 DAs for P-mesons at $\mu^2 = 10~ \text{GeV}^2$.}
\renewcommand{\tabcolsep}{1.5pc}
    \begin{tabular}{cccccc}\hline\hline
    Mesons & ~~I/II &~~$\langle\xi^1\rangle$  & $\langle\xi^2\rangle$ & $\langle\xi^3\rangle$ & $\langle\xi^4\rangle$ \\[4pt]\hline
    \multirow{2}{*}{$D^+$}&I & -0.376 &~~0.244 &~~ -0.159 &~~ 0.117 \\
    &II & -0.375 & ~~0.245 & ~~-0.160 & ~~~0.119 \\
     \cline{1-6}
    \multirow{2}{*}{$D_s$}&I & -0.309 &~~0.198 &~~ -0.115 &~~ 0.081 \\
    &II & -0.329 & ~~0.209 & ~~~-0.125 &~~ 0.089 \\
     \cline{1-6}
    \multirow{2}{*}{$B^+$}&I & -0.660 &~~0.471 &~~ -0.354 &~~ 0.277 \\
    &II & -0.669 & ~~0.482 &~~ -0.365 &~~ 0.286 \\
     \cline{1-6}
    \multirow{2}{*}{$B_s$}&I & -0.618 &~~0.421 &~~ -0.304 &~~ 0.229 \\
    &II & -0.638& ~~0.443 &~~ -0.324 & ~~0.247\\
     \cline{1-6}
    \multirow{2}{*}{$B_c$}&I & -0.373 &~~0.187 &~~ -0.102 &~~ 0.060 \\
    &II & -0.393 & ~~0.196 &~~ -0.106 &~~ 0.062 \\
     \cline{1-6}
     \hline
    \end{tabular}

\end{table}

\begin{table}[!]
  \caption{\label{tab:9}The first four $\xi$-moment $\langle\xi^n\rangle$ of twist-3 DAs for P-mesons at $\mu^2 = 10~ \text{GeV}^2$.}
\renewcommand{\tabcolsep}{1.5pc}
    \begin{tabular}{cccccc}\hline\hline
    Mesons & I/II &~~$\langle\xi^1\rangle^p$  & $\langle\xi^2\rangle^p$ & $\langle\xi^3\rangle^p$ & $\langle\xi^4\rangle^p$ \\[4pt]\hline
    \multirow{2}{*}{$D^+$}&I & -0.471 &~~0.323 &~~ -0.230 &~~ 0.177 \\
    &II & -0.477 & ~~0.331 & ~~-0.238 & ~~~0.186 \\
     \cline{1-6}
    \multirow{2}{*}{$D_s$}&I & -0.373 &~~0.240 &~~ -0.151 &~~ 0.108 \\
    &II & -0.396 & ~~0.256 & ~~~-0.165 &~~ 0.119 \\
     \cline{1-6}
    \multirow{2}{*}{$B^+$}&I &-0.721 &~~0.554 &~~ -0.444 &~~ 0.367 \\
    &II & -0.731 & ~~0.568 &~~ -0.459 &~~ 0.381 \\
     \cline{1-6}
    \multirow{2}{*}{$B_s$}&I & -0.669 &~~0.484 &~~ -0.368 &~~ 0.289 \\
    &II & -0.688 & ~~0.508 &~~ -0.392 & ~~0.312 \\
     \cline{1-6}
    \multirow{2}{*}{$B_c$}&I & -0.399 &~~0.207 &~~ -0.116 &~~ 0.070 \\
    &II & -0.416 & ~~0.214 &~~ -0.119 &~~ 0.071 \\
     \cline{1-6}
     \hline
    \end{tabular}
\end{table}

\begin{table}[!]
  \caption{\label{tab:10}The first four $n$-th transverse moment $\langle \mathbf{k}^n_\perp\rangle$ of twist-2 and twist-3 DAs for P-mesons.}
\renewcommand{\tabcolsep}{0.2pc}
    \begin{tabular}{cccc|cc|cc|cc}\hline\hline
    Mesons & I/II &~~$\langle \mathbf{k}^1_\perp\rangle_{\phi_{2}}$  & $\langle \mathbf{k}^1_\perp\rangle_{\phi_3^P}$ &$\sqrt{\langle \mathbf{k}^2_\perp\rangle_{\phi_{2}}}$  & $\sqrt{\langle \mathbf{k}^2_\perp\rangle_{\phi_3^P}}$ &$\sqrt[3]{\langle \mathbf{k}^3_\perp\rangle_{\phi_{2}}}$  & $\sqrt[3]{\langle \mathbf{k}^3_\perp\rangle_{\phi_3^P}}$ &$\sqrt[4]{\langle \mathbf{k}^4_\perp\rangle_{\phi_{2}}}$  & $\sqrt[4]{\langle \mathbf{k}^4_\perp\rangle_{\phi_3^P}}$\\[4pt]\hline
    \multirow{2}{*}{$D^+$}&I & 0.530 &0.530 &0.602 &0.602&0.666 &0.666&0.724 &0.724 \\
    &II & 0.533 & 0.533 & 0.606 & 0.606& 0.670 & 0.670& 0.729 & 0.729 \\
     \cline{1-10}
    \multirow{2}{*}{$D_s$}&I & 0.596 &0.596& 0.676 &0.676& 0.747 &0.747& 0.812 &0.812 \\
    &II & 0.573 & 0.573& 0.650 & 0.650& 0.718 & 0.718& 0.780 & 0.780 \\
     \cline{1-10}
    \multirow{2}{*}{$B^+$}&I &0.637 &0.637&0.722 &0.722&0.797 &0.797&0.865 &0.865  \\
    &II & 0.627 & 0.627 & 0.712 & 0.712& 0.785 & 0.785& 0.852 & 0.852 \\
     \cline{1-10}
    \multirow{2}{*}{$B_s$}&I & 0.706 &0.706& 0.800 &0.800& 0.883 &0.883& 0.958 &0.958  \\
    &II & 0.680 & 0.680& 0.770 & 0.770& 0.850 & 0.850& 0.922 & 0.922 \\
     \cline{1-10}
    \multirow{2}{*}{$B_c$}&I & 1.057 &1.057& 1.194 &1.194& 1.316 &1.316& 1.425 &1.425  \\
    &II & 0.972& 0.972& 1.097& 1.097& 1.209& 1.209& 1.309& 1.309  \\
     \cline{1-10}
     \hline
    \end{tabular}
\end{table}
Furthermore, we proceed to calculate the DAs of heavy-flavor mesons and present the results in Figs.~\ref{3}-\ref{4} and Tables~\ref{tab:6}-\ref{tab:10}. Some findings are as follows:
\begin{itemize}
\item Figs.~\ref{3}-\ref{4} display the twist-2 and twist-3 DAs of heavy mesons under two schemes. The curves reveal that as the difference between the constituents' masses increases, both the peak value of the DA and its deviation from the symmetric axis at $x=0.5$ become larger. The curves of heavy mesons' DAs exhibit a pattern analogous to Wien's displacement law and we attempt to describe them using an empirical equation $\frac{\Delta m^{0.61}}{\beta^{1.10}}x_p=0.67$, where $\Delta m$ is mass difference between constituents and $x_p$ is the horizontal coordinate corresponding to the peak of the curve. Our previous work has also shown that $\Delta m$  significantly affects the EMFFs and charge radii of mesons~\cite{Xu:2025ntz}. It is easy to find that the peak value of $\phi_2(x)$ is negatively correlated with the most probable fraction $x_p$. With enhanced flavor symmetry breaking, significant differences emerge in the endpoint behavior of the DAs: the contribution near $x=0$ is substantial and thus cannot be neglected in phenomenological calculations and is requires careful treatment. Conversely, the DA vanishes at $x=1$, which indicates a negligible contribution near this endpoint. Particularly, for the $B_c$ meson, the DA reaches zero at both endpoints $x=0,1$ implying a negligible endpoint contributions. The patterns observed for twist-2 and twist-3 DAs are fundamentally consistent.
\item The results of Gegenbauer moments $a_n^{(P)}$ are presented in Tables~\ref{tab:6}-\ref{tab:7} and it is found that for the first Gegenbauer moment $a_1$, our results indicate twist-3 values are approximately twice those of twist-2, $a_1^P\approx 2 a_1$. The more substantial deviation in the peak position suggests that flavor symmetry breaking effects are more significant in higher-twist DAs, which explains the larger values of higher-order Gegenbauer moments $a_n^P$. The values of $\xi$-moments are listed in Table~\ref{tab:8} and the two schemes give comparable results, which is related to $a_n$ by
\begin{eqnarray}
\langle\xi^1\rangle=\frac{3}{5}a_1,~\langle\xi^2\rangle=\frac{12}{35}a_2+\frac{1}{5},~
\langle\xi^3\rangle=\frac{9}{35}a_1+\frac{4}{21}a_3,~
\langle\xi^4\rangle=\frac{3}{35}+\frac{8}{35}a_2+\frac{8}{77}a_4.
\label{eq:27}
\end{eqnarray}
It is easy to check the above relationships. For odd-order cases, for instance $\langle\xi^{1,3}\rangle$, their values are all negative, which is reasonable because $\xi=x-\bar{x}$ and this longitudinal momentum fractions gap tends to be negative as shown in Figs.~\ref{3}-\ref{4}.
\item Table~\ref{tab:10} presents the results for the first four transverse moments $\langle \mathbf{k}^n_\perp\rangle$, which unexpectedly show the identical values for twist-2 and twist-3 DAs. One can check that the $\mathbf{k}^n_\perp$-moments are twist-independent. Interestingly, as the mass of meson (or the parameter $\beta$ of bound state) increases, the value of $\sqrt[n]{\langle \mathbf{k}^n_\perp\rangle}$ also gradually increases, which indicates that the parameter $\beta$ contains implicitly information about transverse scale of the bound state. Meanwhile, $\sqrt[n]{\langle \mathbf{k}^n_\perp\rangle}$ increases almost in equal value as $n$ increases.
\end{itemize}
\section{CONCLUSIONS}\label{sec:4}

The twist-2 and twist-3 DAs for pseudoscalar mesons are investigated within LFQM in this work. The normalization condition for twist-3 DA $\phi^P_3(x)$ is remarkably improved by the factor $\mu_M=M_0^2/(m_q+m_{\bar{q}})$ with the replacement $M\to M_0$. The numerical results are carried out with two schemes, where the parameter sets are fixed by the constraints from the mesonic decay constants and masses spectrum, respectively. In addition, the Gegenbauer moment $a_n$, $\xi$-moment $\langle \xi^n\rangle$ and transverse moment $\langle \mathbf{k}^n_\perp\rangle$ have also been analysed, in which the findings are as follows:

For pion, the Gegenbauer moment $a_{2,\pi}=0.054$ under scheme-I well consists with predictions of the  platykurtic model and NL$\chi$QM, while the $a_{2,\pi}=0.126$ under scheme-II aligns well with outcomes from the LQCD, the DSE and data-driven LCSR analyses. The results of $\phi_{2,\pi}(x)$ and $\phi^P_{3,\pi}(x)$ are identical with the replacement $M\to M_0$. For kaon, the flavor symmetry breaking effects are more significant in higher-twist DAs, which also holds in heavy meson systems. It is found that the first Gegenbauer moment of twist-3 DA are approximately twice those of twist-2 DA for heavy mesons, i.e.,  $a_1^P\approx 2 a_1$. Interestingly, the numerical analysis reveals that transverse moment $\langle \mathbf{k}^n_\perp\rangle$ is twist-independent with the replacement $M\to M_0$. In addition, an empirical equation $\frac{\Delta m^{0.61}}{\beta^{1.10}}x_p=0.67$ is suggested to describe the curves of heavy mesons' DAs.
Collectively, these results demonstrate that mass asymmetry between the constituent quarks profoundly influences the shape of DAs and their associated moments, which confirms the findings in research of EMFFs. These theoretical predictions are expected to be tested in future high-energy experiments.

\section*{Acknowledgements}

We are grateful to Xing-Gang Wu for very valuable discussions. Qin Chang is supported by the National Natural Science Foundation of China (Grant No.12275067), Science and Technology R$\&$D Program Joint Fund Project of Henan Province  (Grant No.225200810030), Science and Technology Innovation Leading Talent Support Program of Henan Province (Grant No.254200510039), and National Key R$\&$D Program of China (Grant No.2023YFA1606000). Xiao-Nan Li is supported by the Tongling University Talent Program (Grant No.R23100).

\section{Appendix: The details of calculations for LF matrix elements}\label{sec:app}
We present the details of calculations for LF matrix elements $\mathcal{M}_{\Gamma}\equiv\langle 0|\bar{q}(z)\Gamma q(-z)|M(P)\rangle$ step by step in this appendix. At leading-order eigenstates of meson, the bound state can be expressed as Eq.\ref{eq:1}
\begin{eqnarray}
|M(P)\rangle = \int \{d^3p_q\}\{d^3p_{\bar{q}}\}2(2\pi)^3\delta^3(P-p_q-p_{\bar{q}}) \sum_{h,\bar{h}}\Psi_{h\bar{h}}(x,\mathbf{k}_\perp)|q(p_q,h)\bar{q}(p_{\bar{q}},\bar{h})\rangle,
\label{eq:app1}
\end{eqnarray}
inwhich the Dirac's ket $|q(p_q,h)\bar{q}(p_{\bar{q}},\bar{h})\rangle$ can be further expressed as
\begin{eqnarray}
|q(p_q,h)\bar{q}(p_{\bar{q}},\bar{h})\rangle=\sqrt{2p_q^+}\sqrt{2p_{\bar{q}}^+}b_h^\dag(p_q)d_{\bar{h}}^\dag(p_{\bar{q}})|0\rangle.
\label{eq:app2}
\end{eqnarray}
The quark and antiquark field are given as
\begin{eqnarray}
q(-z)=\int\frac{dp_q^+}{\sqrt{2p_q^+}}\frac{d^2\mathbf{p}_{q\perp}}{(2\pi)^3}\sum_h[b_h(p_q^+,\mathbf{p}_{q\perp})u_h(p_q^+,\mathbf{p}_{q\perp})
e^{ip_q\cdot z}+d_h^\dagger(p_q^+,\mathbf{p}_{q\perp})\nu_h(p_q^+,\mathbf{p}_{q\perp})e^{-ip_q\cdot z}],\\
\bar{q}(z)=\int\frac{dp_{\bar{q}}^+}{\sqrt{2p_{\bar{q}}^+}}\frac{d^2\mathbf{p}_{\bar{q}\perp}}{(2\pi)^3}\sum_{\bar{h}}[b^\dag_{\bar{h}}(p_{\bar{q}}^+,\mathbf{p}_{\bar{q}\perp})\bar{u}_{\bar{h}}(p_{\bar{q}}^+,\mathbf{p}_{\bar{q}\perp})
e^{ip_{\bar{q}}\cdot z}+d_{\bar{h}}(p_{\bar{q}}^+,\mathbf{p}_{\bar{q}\perp})\bar{\nu}_{\bar{h}}(p_{\bar{q}}^+,\mathbf{p}_{\bar{q}\perp})e^{-ip_{\bar{q}}\cdot z}].
\label{eq:app3}
\end{eqnarray}
Applying the Eqs.~\ref{eq:app2}-\ref{eq:app3} to the matrix element $\mathcal{M}_{\Gamma}$, we can get totally four terms inwhich there is just one term survival as $\langle0|d_{\bar{h}}b_hb_h^\dag d_{\bar{h}}^\dag[\cdot\cdot\cdot]|0\rangle$. For simplicity, the retained quantities are marked in red in Eq.~\ref{eq:app3}. With the identity $p_q\cdot z-p_{\bar{q}}\cdot z=(p_q-p_{\bar{q}})\cdot z=(x-\bar{x})P\cdot z$, we get the general form for the LF matrix elements
\begin{eqnarray}
\mathcal{M}_{\Gamma}=\sqrt{N_c}\sum_{h,\bar{h}}\int\frac{dp_q^+}{\sqrt{2p_q^+}}\frac{d^2\mathbf{p}_\perp}{(2\pi)^3}
\Psi_{h,\bar{h}}(x,\mathbf{k}_\perp)\bar{\nu}_{\bar{h}}\Gamma u_h e^{i(x-\bar{x})P\cdot z}.
\label{eq:app4}
\end{eqnarray}



\begin{thebibliography}{99}
\bibitem{Lepage:1980fj}
G.~P.~Lepage and S.~J.~Brodsky,
Phys. Rev. D \textbf{22}, 2157 (1980)

\bibitem{Brodsky:1997de}
S.~J.~Brodsky, H.~C.~Pauli and S.~S.~Pinsky,
Phys. Rept. \textbf{301}, 299-486 (1998)

\bibitem{Efremov:1979qk}
A.~V.~Efremov and A.~V.~Radyushkin,
Phys. Lett. B \textbf{94}, 245-250 (1980)

\bibitem{Collins:1981uw}
J.~C.~Collins and D.~E.~Soper,
Nucl. Phys. B \textbf{194}, 445-492 (1982)

\bibitem{Braun:1989iv}
V.~M.~Braun and I.~E.~Filyanov,
Z. Phys. C \textbf{48}, 239-248 (1990)
\bibitem{Jaus:1989au}
W.~Jaus,
Phys. Rev. D \textbf{41}, 3394 (1990)

\bibitem{Jaus:1989av}
W.~Jaus and D.~Wyler,
Phys. Rev. D \textbf{41}, 3405 (1990)

\bibitem{Cheng:1996if}
H.~Y.~Cheng, C.~Y.~Cheung and C.~W.~Hwang,
Phys. Rev. D \textbf{55}, 1559-1577 (1997)

\bibitem{ODonnell:1996sya}
P.~J.~O'Donnell and G.~Turan,
Phys. Rev. D \textbf{56}, 295-302 (1997)

\bibitem{Cheung:1996qt}
C.~Y.~Cheung, C.~W.~Hwang and W.~M.~Zhang,
Z. Phys. C \textbf{75}, 657-664 (1997)

\bibitem{Choi:1996mq}
H.~M.~Choi and C.~R.~Ji,
Nucl. Phys. A \textbf{618}, 291-316 (1997)

\bibitem{Barik:1997qq}
N.~Barik, S.~K.~Tripathy, S.~Kar and P.~C.~Dash,
Phys. Rev. D \textbf{56}, 4238-4249 (1997)

\bibitem{Hwang:2000ez}
C.~W.~Hwang,
Eur. Phys. J. C \textbf{19}, 105-111 (2001)

\bibitem{Hwang:2010hw}
C.~W.~Hwang,
Phys. Rev. D \textbf{81}, 114024 (2010)

\bibitem{Geng:2001de}
C.~Q.~Geng, C.~W.~Hwang, C.~C.~Lih and W.~M.~Zhang,
Phys. Rev. D \textbf{64}, 114024 (2001)

\bibitem{Chang:2018aut}
Q.~Chang, X.~N.~Li, X.~Q.~Li and F.~Su,
Chin. Phys. C \textbf{42}, no.7, 073102 (2018)
\bibitem{Chang:2016ouf}
Q.~Chang, S.~J.~Brodsky and X.~Q.~Li,
Phys. Rev. D \textbf{95}, no.9, 094025 (2017)

\bibitem{Chang:2018mva}
Q.~Chang, L.~L.~Chen and S.~Xu,
J. Phys. G \textbf{45}, no.7, 075005 (2018)

\bibitem{Chang:2019obq}
Q.~Chang, L.~T.~Wang and X.~N.~Li,
JHEP \textbf{12}, 102 (2019)

\bibitem{Chang:2019mmh}
Q.~Chang, X.~N.~Li and L.~T.~Wang,
Eur. Phys. J. C \textbf{79}, no.5, 422 (2019)

\bibitem{Choi:2017uos}
H.~M.~Choi and C.~R.~Ji,
Phys. Rev. D \textbf{95}, no.5, 056002 (2017)
\bibitem{Efremov:2009dx}
A.~Efremov and A.~Radyushkin,
Mod. Phys. Lett. A \textbf{24}, 2803-2824 (2009)

\bibitem{Collins:1996fb}
J.~C.~Collins, L.~Frankfurt and M.~Strikman,
Phys. Rev. D \textbf{56}, 2982-3006 (1997)
\bibitem{Dong:2008zza}
Y.~B.~Dong,
Phys. Rev. C \textbf{77}, 015201 (2008)

\bibitem{LatticeParton:2022zqc}
J.~Hua \textit{et al.} [Lattice Parton],
Phys. Rev. Lett. \textbf{129}, no.13, 132001 (2022)

\bibitem{Chen:2023byr}
L.~B.~Chen, W.~Chen, F.~Feng and Y.~Jia,
Phys. Rev. Lett. \textbf{132}, no.20, 201901 (2024)

\bibitem{RQCD:2019osh}
G.~S.~Bali \textit{et al.} [RQCD],
JHEP \textbf{08}, 065 (2019)


\bibitem{Chernyak:1983ej}
V.~L.~Chernyak and A.~R.~Zhitnitsky,
Phys. Rept. \textbf{112}, 173 (1984)

\bibitem{Huang:2013yya}
T.~Huang, T.~Zhong and X.~G.~Wu,
Phys. Rev. D \textbf{88}, 034013 (2013)

\bibitem{Khodjamirian:2004ga}
A.~Khodjamirian, T.~Mannel and M.~Melcher,
Phys. Rev. D \textbf{70}, 094002 (2004)

\bibitem{Ball:2006wn}
P.~Ball, V.~M.~Braun and A.~Lenz,
JHEP \textbf{05}, 004 (2006)

\bibitem{Guo:1991eb}
X.~H.~Guo and T.~Huang,
Phys. Rev. D \textbf{43}, 2931-2938 (1991)

\bibitem{Chai:2025xuz}
J.~Chai and S.~Cheng,
JHEP \textbf{06}, 229 (2025)

\bibitem{Cheng:2019ruz}
S.~Cheng,
Phys. Rev. D \textbf{100}, no.1, 013007 (2019)
\bibitem{Chang:2013pq}
L.~Chang, I.~C.~Cloet, J.~J.~Cobos-Martinez, C.~D.~Roberts, S.~M.~Schmidt and P.~C.~Tandy,
Phys. Rev. Lett. \textbf{110}, no.13, 132001 (2013)

\bibitem{Raya:2015gva}
K.~Raya, L.~Chang, A.~Bashir, J.~J.~Cobos-Martinez, L.~X.~Guti{\'e}rrez-Guerrero, C.~D.~Roberts and P.~C.~Tandy,
Phys. Rev. D \textbf{93}, no.7, 074017 (2016)

\bibitem{Chang:2013epa}
L.~Chang, C.~D.~Roberts and S.~M.~Schmidt,
Phys. Lett. B \textbf{727}, 255-259 (2013)

\bibitem{Shi:2015esa}
C.~Shi, C.~Chen, L.~Chang, C.~D.~Roberts, S.~M.~Schmidt and H.~S.~Zong,
Phys. Rev. D \textbf{92}, 014035 (2015)

\bibitem{Roberts:2021nhw}
C.~D.~Roberts, D.~G.~Richards, T.~Horn and L.~Chang,
Prog. Part. Nucl. Phys. \textbf{120}, 103883 (2021)

\bibitem{Nam:2006sx}
S.~i.~Nam and H.~C.~Kim,
Phys. Rev. D \textbf{74}, 076005 (2006)

\bibitem{Nam:2006au}
S.~i.~Nam, H.~C.~Kim, A.~Hosaka and M.~M.~Musakhanov,
Phys. Rev. D \textbf{74}, 014019 (2006)
\bibitem{Agaev:2010aq}
S.~S.~Agaev, V.~M.~Braun, N.~Offen and F.~A.~Porkert,
Phys. Rev. D \textbf{83}, 054020 (2011)


\bibitem{Agaev:2012tm}
S.~S.~Agaev, V.~M.~Braun, N.~Offen and F.~A.~Porkert,
Phys. Rev. D \textbf{86}, 077504 (2012)

\bibitem{Cheng:2020vwr}
S.~Cheng, A.~Khodjamirian and A.~V.~Rusov,
Phys. Rev. D \textbf{102}, no.7, 074022 (2020)

\bibitem{Chai:2022srx}
J.~Chai, S.~Cheng and J.~Hua,
Eur. Phys. J. C \textbf{83}, no.7, 556 (2023)

\bibitem{RuizArriola:2002bp}
E.~Ruiz Arriola and W.~Broniowski,
Phys. Rev. D \textbf{66}, 094016 (2002)

\bibitem{Praszalowicz:2001wy}
M.~Praszalowicz and A.~Rostworowski,
Phys. Rev. D \textbf{64}, 074003 (2001)

\bibitem{Petrov:1998kg}
V.~Y.~Petrov, M.~V.~Polyakov, R.~Ruskov, C.~Weiss and K.~Goeke,
Phys. Rev. D \textbf{59}, 114018 (1999)

\bibitem{Nam:2006mb}
S.~i.~Nam and H.~C.~Kim,
Phys. Rev. D \textbf{74}, 096007 (2006)


\bibitem{Choi:2009ai}
H.~M.~Choi and C.~R.~Ji,
Phys. Rev. D \textbf{80}, 054016 (2009)

\bibitem{Zhao:2018zcb}
Z.~X.~Zhao,
Chin. Phys. C \textbf{42}, no.9, 093101 (2018)

\bibitem{Wuenqi:2025tup}
Wuenqi, R.~H.~Li and Z.~X.~Zhao,
Eur. Phys. J. C \textbf{85}, no.9, 1023 (2025)

\bibitem{Xu:2025aow}
S.~Xu, X.~N.~Li and X.~G.~Wu,
Chin. Phys. Lett. \textbf{42}, no.8, 080201 (2025)

\bibitem{Ke:2010vn}
H.~W.~Ke, X.~Q.~Li, Z.~T.~Wei and X.~Liu,
Phys. Rev. D \textbf{82}, 034023 (2010)

\bibitem{Xu:2025ntz}
S.~Xu, X.~N.~Li and X.~G.~Wu,
[arXiv:2507.07523 [hep-ph]].
\bibitem{Dirac:1949cp}
P.~A.~M.~Dirac,
Rev. Mod. Phys. \textbf{21}, 392-399 (1949)

\bibitem{Brodsky:1998hn}
S.~J.~Brodsky and D.~S.~Hwang,
Nucl. Phys. B \textbf{543}, 239-252 (1999)

\bibitem{Keister:1991sb}
B.~D.~Keister and W.~N.~Polyzou,
Adv. Nucl. Phys. \textbf{20}, 225-479 (1991)

\bibitem{Coester:1992cg}
F.~Coester,
Prog. Part. Nucl. Phys. \textbf{29}, 1-32 (1992)

\bibitem{Szczepaniak:1995vn}
A.~Szczepaniak, C.~R.~Ji and S.~R.~Cotanch,
Phys. Rev. D \textbf{52}, 5284-5294 (1995)

\bibitem{Osborn:1972dy}
H.~Osborn,
Nucl. Phys. B \textbf{38}, 429-465 (1972)

\bibitem{Susskind:1967rg}
L.~Susskind,
Phys. Rev. \textbf{165}, 1535-1546 (1968)

\bibitem{Bardakci:1968zqb}
K.~Bardakci and M.~B.~Halpern,
Phys. Rev. \textbf{176}, 1686-1699 (1968)

\bibitem{Chang:1968bh}
S.~J.~Chang and S.~K.~Ma,
Phys. Rev. \textbf{180}, 1506-1513 (1969)

\bibitem{Brodsky:1973kb}
S.~J.~Brodsky, R.~Roskies and R.~Suaya,
Phys. Rev. D \textbf{8}, 4574 (1973)

\bibitem{Bjorken:1970ah}
J.~D.~Bjorken, J.~B.~Kogut and D.~E.~Soper,
Phys. Rev. D \textbf{3}, 1382 (1971)

\bibitem{Neville:1971uc}
R.~A.~Neville and F.~Rohrlich,
Phys. Rev. D \textbf{3}, 1692-1707 (1971)
\bibitem{Brodsky:2014yha}
S.~J.~Brodsky, G.~F.~de Teramond, H.~G.~Dosch and J.~Erlich,
Phys. Rept. \textbf{584}, 1-105 (2015)

\bibitem{Melosh:1974cu}
H.~J.~Melosh,
Phys. Rev. D \textbf{9}, 1095 (1974)


\bibitem{ParticleDataGroup:2022pth}
R.~L.~Workman \textit{et al.} [Particle Data Group],
PTEP \textbf{2022}, 083C01 (2022)




\bibitem{Zhong:2014jla}
T.~Zhong, X.~G.~Wu, Z.~G.~Wang, T.~Huang, H.~B.~Fu and H.~Y.~Han,
Phys. Rev. D \textbf{90}, no.1, 016004 (2014)

\bibitem{Choi:2007yu}
H.~M.~Choi and C.~R.~Ji,
Phys. Rev. D \textbf{75}, 034019 (2007)

\bibitem{Li:1992nu}
H.~n.~Li and G.~F.~Sterman,
Nucl. Phys. B \textbf{381}, 129-140 (1992)

\bibitem{Li:1992ce}
H.~n.~Li,
Phys. Rev. D \textbf{48}, 4243-4254 (1993)

\bibitem{Stefanis:2020rnd}
N.~G.~Stefanis,
Phys. Rev. D \textbf{102}, no.3, 034022 (2020)

\end{thebibliography}
\end{document}